**YingLong-weather: AI-Based Limited Area Models for Forecasting**

**of Non-precipitation Surface Meteorological Variables**

Pengbo Xu,[1,2] Xiaogu Zheng,[2,7] Tianyan Gao,[3] Yu Wang,[1] Junping Yin,[2,4,5,✉] Juan Zhang,[2,6] Xuanze Zhang,[8] San Luo,[9] Zhonglei Wang,[10] Zhimin Zhang,[11] Xiaoguang Hu,[11] and Xiaoxu Chen[11]

[1] *School of Mathematical Sciences, Key Laboratory of MEA (Ministry of Education), Shanghai Key Laboratory of PMMP, East China Normal University, Shanghai 200241, P.R. China.*

[2] *Shanghai Zhangjiang Institute of Mathematics, Shanghai, 201203, P.R. China.*

[3] *School of Mathematics and Statistics, Northeast Normal University, Changchun 130000, P.R. China.*

[4] *Institute of Applied Physics and Computational Mathematics, Beijing, 100094, China.*

[5] *National Key Laboratory of Computational Physics, Beijing 100088, China.*

[6] *Institute of Artificial Intelligence, Beihang University, Beijing, 100191, China.*

[7] *International Global Change Institute, Hamilton, New Zealand.*

[8] *Key Laboratory of Water Cycle and Related Land Surface Processes, Institute of Geographic Sciences and Natural Resources Research, Chinese Academy of Sciences, Beijing 100101, China.*

[9] *China Meteorological Administration Basin Heavy Rainfall Key Laboratory/Hubei Key Laboratory for Heavy Rain Monitoring and Warning Research, Institute of Heavy Rain, China Meteorological Administration, Wuhan 430205, China.*

[10] *Wang Yanan Institute for Studies in Economics and School of Economics, Xiamen University, Xiamen, Fujian, 361005, China.*

[11] *Baidu Inc., Beijing, China, 100085.*

✉ *Corresponding author, yinjp829829@126.com*

## Abstract

Recently, artificial intelligence-based (AI-based) models for forecasting of global weather have been rapidly developed. Most of the global models are trained on reanalysis datasets with a spatial resolution of 0.25° ×0.25°. However, research on AI-based high spatial resolution limited area weather forecasting models remains limited. In this study, YingLong, an AI-based limited area weather forecasting model with a spatial resolution of 3 km × 3 km is developed. YingLong employs a parallel structure of global and local blocks to capture multiscale meteorological features and operates much faster than the dynamical limited area model WRF-ARW. In two selected limited areas (one relatively flat and the other featuring significant mountain ranges), YingLong (with lateral boundary condition imposed by the global AI-based model Pangu-weather) demonstrates superior skill in forecasting surface wind speed compared to WRF-ARW. Additionally, it shows comparable skill in forecasting surface temperature and pressure. The accuracy of surface temperature and humidity forecasts can be further improved by applying better boundary conditions. YingLong also addresses issues related to the lateral boundary conditions of AI-based limited area models, such as selecting the width of the lateral boundary region and combining finer and coarser resolution predictions in this region. Therefore, YingLong has a great potential to generate cost-effective multiyear high-resolution synthetic wind speed that maintain meteorological realism both spatially and temporally, aiding in the planning and operations for wind power generation companies.

## 1 Introduction

Accurate weather forecasting with higher resolution plays a crucial role in modern society, especially for surface meteorological variables [1]. Research in numerical weather prediction (NWP) has advanced rapidly over the past decades [2, 3]. Traditional NWP involves numerically solving the governing equations of fluid dynamics [4, 5], which is often computationally expensive and time-consuming [6-10].

Recently, several AI-based global weather forecasting models have been developed, such as FourCastNet [21], Pangu-Weather [22], GraphCast [23], FengWu [24], FuXi [25], ClimaX [26], Aurora [27], and AIFS [28]. Most of these models are trained using global ERA5 reanalysis data [29] with a spatial resolution of $0.25° \times 0.25°$, matching the ECMWF Integrated Forecasting System [30]. AI-based models consume significantly fewer computational resources and run much faster than NWPs. Since ERA5 reanalysis data is generated by assimilating observations into NWP forecasts, it can correct some model errors related to NWP, such as errors associated with the parameterization of physical processes [31]. Therefore, AI-based weather forecasting models trained on ERA5 reanalysis data could potentially be more accurate than traditional NWPs.

For weather forecasts with finer spatial resolution (such as 3 km) in domains ranging from several countries to a county or city, limited area models (LAMs) are developed. However, most work on LAMs is related to NWP, and studies on AI-based LAMs remain limited. The MetNet series [32, 33] is an attempt in this area, pre-trained using finer spatial resolution analysis data and focusing on precipitation, radar, satellite, and ground site observations. Based on the architecture of GraphCast, Oskarsson et al. [34] made forecasts in the Nordic region by completely replacing the lateral boundary region with NWP data at the corresponding time.

Recently, AI-based models with high resolution in a region have been established by training on both fine resolution regional analysis data and coarse resolution global reanalysis data. For example, Pathak et al. [35] applied generative diffusion modeling trained with 3 km resolution analysis data from the High-Resolution Rapid Refresh (HRRR) and 0.25° resolution ERA5 data for high-resolution forecasts in Central America. Nipen et al. [36] used a graph neural network trained with 2.5 km resolution operational analyses from the MetCoOp Ensemble Prediction System data [37] and ERA5 data for high-resolution forecasts in the Nordics. Recently, Zhao et al [38] used observation data from the past 6 hours from meteorological stations and satellites to train an AI-based LAM for forecasting non-precipitation surface meteorological variables on a resolution of 0.05°×0.05°. However, due to the lack of imposed lateral boundary conditions (LBC), only 12-hour effective forecasts are available.

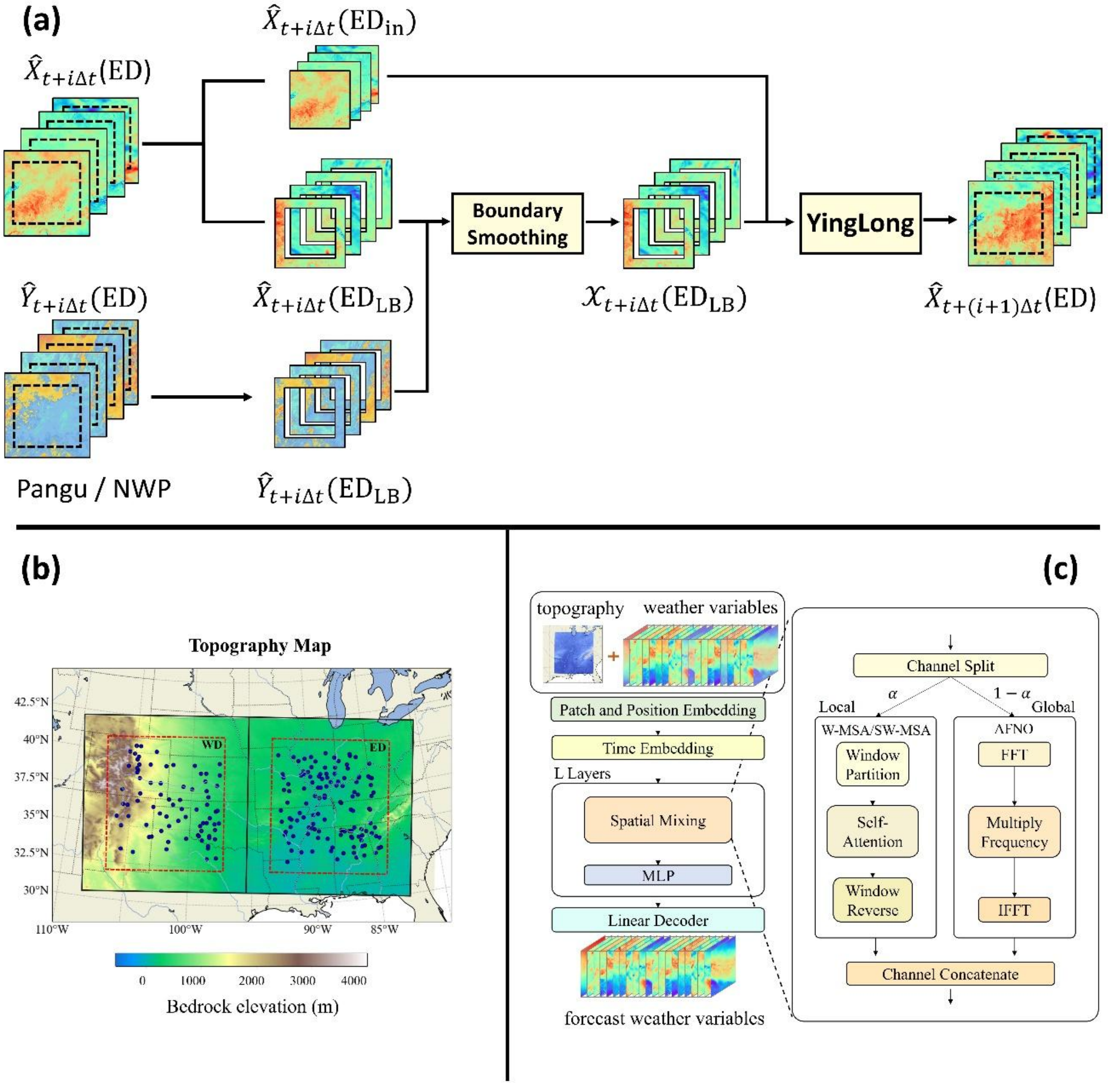

Figure 1. YingLong model framework and regional topography. a) Rolling forecasts of YingLong on the region ED through the process in Sec. 4.5: The prediction of Yinglong on the ED $\hat{X}_{t+i\Delta t}(\mathrm{ED})$ is divided into $\hat{X}_{t+i\Delta t}(\mathrm{ED_{in}})$ on the inner area and $\hat{X}_{t+i\Delta t}(\mathrm{ED_{LB}})$ on the lateral boundary region. Then boundary smoothing (Eq. (3)) is applied to $\hat{X}_{t+i\Delta t}(\mathrm{ED_{LB}})$ and $\hat{Y}_{t+i\Delta t}(\mathrm{ED_{LB}})$ (the forecast of Pangu on $\mathrm{ED_{LB}}$) to form $\mathcal{X}_t(\mathrm{ED_{LB}})$. Finally, $\hat{X}_{t+i\Delta t}(\mathrm{ED_{in}})$ and $\mathcal{X}_t(\mathrm{ED_{LB}})$ are merged for the initial value of YingLong prediction $\hat{X}_{t+(i+1)\Delta t}(\mathrm{ED})$. b) The domains ED and WD. The dots represent observation sites. c) The architecture of YingLong.

In this study, an AI-based LAM called YingLong is developed for forecasting non-precipitation surface meteorological variables, including surface temperature, pressure, humidity, and wind speed. These variables are highly sensitive to human life, alongside precipitation. Unlike previous work, YingLong is trained solely using 3 km resolution HRRR regional analysis data. Although global coarse resolution data are not used for training, they are required to supply the LBC. This study investigates YingLong with LBC imposed by Pangu-weather for forecasting non-precipitation surface meteorological variables.

YingLong is also applied to investigate issues related to LBCs of AI-based LAMs. As is well known, the domain of a LAM is divided into an inner area and a surrounding lateral boundary region. The meteorological conditions on the lateral boundary region are provided by coarse-resolution models, while the inner area is the focus of research and application. The coarse-resolution atmospheric information on the lateral boundary region is transmitted to the inner area through the forecast of the high-resolution LAM. Numerical LAMs require extensive processing of coarse resolution LBCs to accommodate differences in model structure and parameterization schemes, ensuring that certain physical constraints are met. However, this may not be necessary for AI-based LAMs. As statistical models, AI-based LAMs have fewer physical restrictions on model structure and initial values compared to numerical models. Nevertheless, effectively transferring meteorological information from the lateral boundary region to the inner area is also a key objective of this paper.

Table 1. The weather variables of YingLong include a total of 24 variables. These consist of 20 upper-air variables from 4 pressure levels (50 hPa, 500 hPa, 850 hPa, and 1000 hPa, following FourCastNet) and four surface variables.

| Type | Full name | Abbreviation |
|---|---|---|
| Upper air variable | geopotential height | Z |
| | specific humidity | S |
| | temperature | T |
| | U component of wind speed | U |
| | V component of wind speed | V |
| Surface variables | Mean sea level pressure | MSLP |
| | 2m temperature | T2M |
| | U component of 10m wind speed | U10 |
| | V component of 10m wind speed | V10 |

## 2. Results

This section is dedicated to the construction of YingLong and its predictability with LBC imposed by Pangu-weather for the two selected limited areas in North America.

### 2.1 Construction of YingLong

YingLong is an AI-based limited area weather forecasting model with a 3 km spatial resolution and 24 weather variables (see Table 1). It is trained on hourly HRRR analysis data (created by assimilating observations into WRF-ARW) in the period 2015-2021, and validated and tested on HRRR analysis data in 2023 and 2022, respectively (see Methods 4.1 for more details). To test its sensitivity to topography, YingLong is constructed and evaluated on two domains: the east domain (ED), which is relatively flat, and the west domain (WD), which includes considerable mountain ranges (see Fig. 1(b)). Both ED and WD are divided into an inner area (the area within the red dashed line) and a lateral boundary region (the region outside the red dashed line).

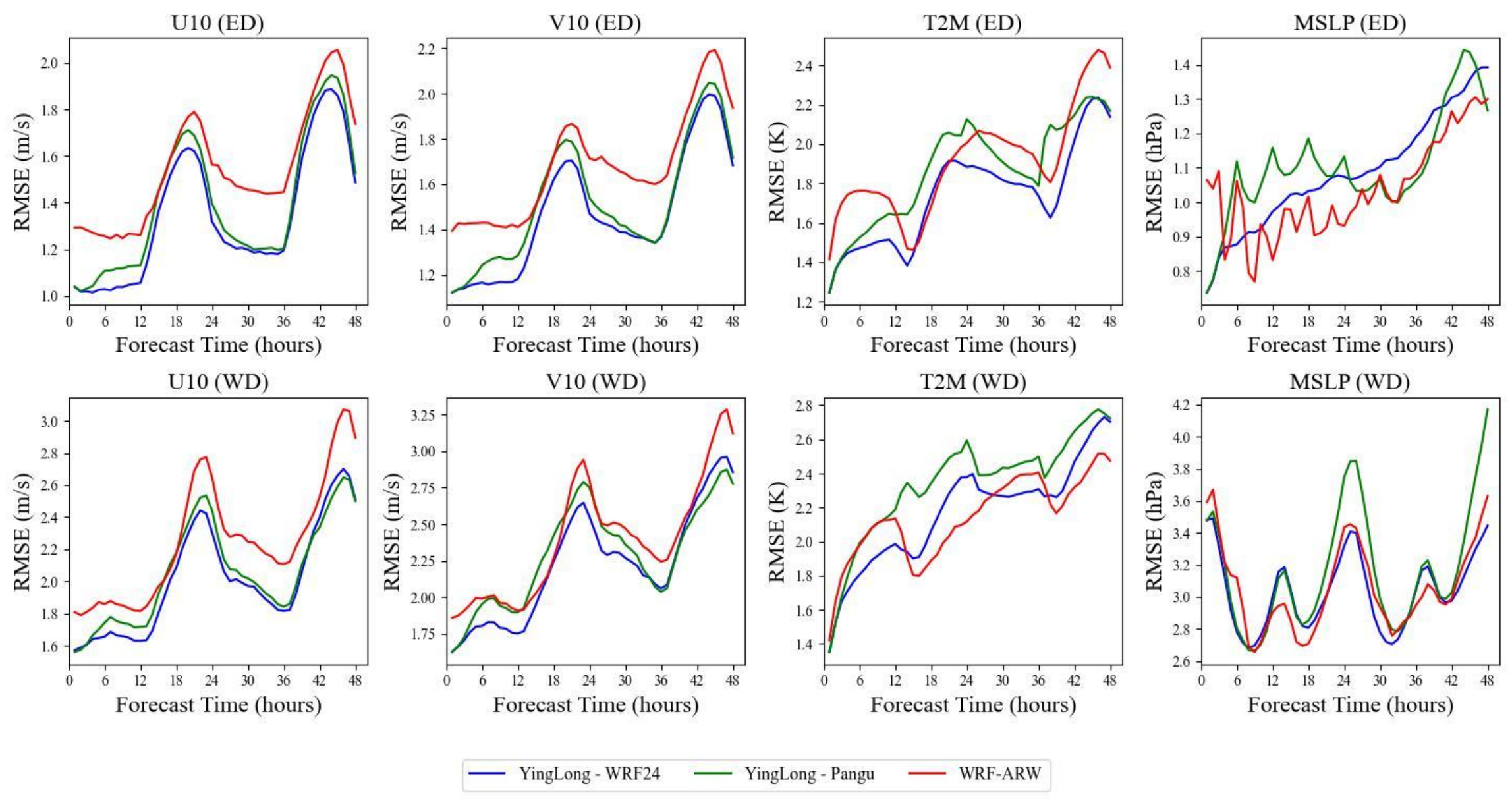

Figure 2. RMSEs with *in-situ* observations as benchmark of WRF-ARW (red lines), YingLong-WRF24 (blue lines), and YingLong-Pangu (green lines) in ED and WD.

The architecture of YingLong includes the embedding layer, the spatial mixing layer (comprising the "Local" branch and "Global" branch), and the linear decoder (see Fig. 1(c) and Methods 4.2 for details). YingLong is trained using the procedure documented in Methods 4.3, with a smoothing boundary condition imposed (see Methods 4.4). The procedure for YingLong's rolling forecasting is documented in Methods 4.5 and briefly illustrated in Fig. 1(a). Two indicators, root mean squared error (RMSE, Eq. (S1) in Sec. SI 1 of Supplementary Material) and anomaly correlation coefficient (ACC, Eq. (S2) in Sec. SI 1 of Supplementary Material), are used to evaluate the models. Specifically, a smaller RMSE or a larger ACC indicates a better performing model.

**2.2 Forecasting skills of YingLong-Pangu**

To construct a fully AI-based LAM with fine spatial resolution, the Pangu-weather medium-term forecasts are applied as the LBC of YingLong (denoted as YingLong-Pangu). To evaluate its forecasting skills more objectively, the *in-situ* observation data [39] is used as the benchmark to calculate the RMSEs of the forecasts. In this study, the forecast for a station observation using a model with 3 km spatial resolution is defined as the model forecast at the 3 km grid containing the observation site.

Figure 2 shows the RMSEs of 48-hour rolling forecasts for YingLong-Pangu and WRF-ARW (Advanced Research WRF [40]), starting at 00:00 UTC, for non-precipitation surface variables (U10, V10, T2M, MSLP) at the stations in the inner areas of ED and WD in 2022. The results indicate that the RMSEs of YingLong-Pangu (green lines) are generally smaller than those of WRF-ARW (red lines) for U10 and V10. However, this is not the case for T2M and MSLP. For T2M in ED, the RMSEs of YingLong-Pangu and WRF-ARW are comparable, while for T2M in WD and MSLP in both domains, the RMSEs of YingLong-Pangu are generally larger than those of WRF-ARW.

To test the impact of the LBC provided by the Pangu-weather forecast on the performance of YingLong, the WRF-ARW forecasts with 24 km spatial resolution are applied as the LBC for YingLong (denoted as YingLong-WRF24). Fig. 2 shows that the RMSEs of YingLong-WRF24 (blue lines) are generally smaller than those of YingLong-Pangu for all non-precipitation surface variables. Moreover, the RMSEs of YingLong-WRF24 are comparable to those of WRF-ARW for T2M and MSLP in WD, and even smaller than those of WRF-ARW for T2M in ED. These results indicate that the WRF-ARW forecasts with 24 km spatial resolution provide a better LBC than the Pangu-weather forecasts. This suggests that the performance of YingLong can be further improved by providing better LBCs.

The conclusions drawn from the RMSEs in Fig. 2 are consistent with those from the ACCs (Fig. S4 in Supplementary Material). These conclusions are further confirmed by the selected visualizations of YingLong-Pangu forecasts (Sec. SI 7 of Supplementary Material). Although surface specific humidity observations are available, it is not a variable included in YingLong (see Table 1). Therefore, there are no rolling forecasts for surface specific humidity observations.

The skills of both YingLong-Pangu and YingLong-WRF24 are higher than those of WRF-ARW for predicting U10 and V10, while this is not always the case for predicting T2M and SLP. The reason could be that U10 and V10 are hydrodynamic variables, while T2M and MSLP are thermodynamic variables that are more sensitive to regional surface conditions, which are not currently described in detail in AI-based models. Specifically, U10 and V10 relate to fluids in motion, which are dominated by pressure gradients. The AI-based YingLong is trained using the analysis states of pressure at several geopotential heights (see Table 1). Since the analysis was generated by assimilating observations into WRF-ARW forecasts, some model errors related to WRF-ARW could be corrected by the observations. Therefore, U10 and V10 are likely better predicted by YingLong than by WRF-ARW. On the other hand, T2M and MSLP are more closely related to surface radiative forcing. The radiative forcing data is not used for training YingLong, but it is a forcing variable for WRF-ARW. Therefore, T2M and MSLP could be better predicted by WRF-ARW than by YingLong. The RMSEs (and ACCs) of all forecasts in ED are larger (smaller) than those in WD. This could be due to the more sophisticated terrain in WD compared to ED.

Accurate forecasting of wind speed is beneficial for renewable energy filed. The wind speed at wind turbine levels (i.e., 85 m) can be estimated from U10 and V10 using a power law, with the power parameter estimated using high-resolution analysis data or numerical LAM simulations (see Eq. (1) of [41]). Wind data at a 1-hour time scale is a crucial input for modeling the performance and impact of wind farms on many countries' electricity systems [41]. Since YingLong can forecast hourly high-resolution wind surface speed more promptly and accurately than NWPs, it has the potential to be applied to the planning and operation of the wind power generation industry.

## 3. Discussions

Results 2.2 relate to the forecast skill of YingLong-Pangu, with *in-situ* observation data used as the benchmark. This section addresses modeling issues for improving YingLong, such as the architecture, the LBCs, the homogeneity of training data, and the extremes simulated by YingLong.

The HRRR analysis is used as the benchmark for the evaluation and testing of YingLong because it serves as the labels for training YingLong. If other data were used as a benchmark, it would be difficult to determine whether the improvement of a model is due to the model itself or because the benchmark fits the model better. Since the HRRR analysis data is a fusion of WRF-ARW and observations, WRF-ARW forecasts with 24 km resolution are chosen as LBC, because they are likely to be more consistent with YingLong forecasts than other LBCs studied in this paper. Therefore, the modeling issues are less affected by the choice of LBCs. Moreover, the smooth LBC scheme (see Eq. (3)) is imposed, and 207km is selected as the width of the lateral boundary region. These choices are confirmed as good in Sec. 3.2.

### 3.1 AI architecture

To select a suitable deep neural network architecture, a comprehensive ablation experiment was conducted to compare different model architectures, including the adaptive Fourier neural operator (AFNO [42], the main module of FourCastNet [21]), the SWIN transformer (the main module of Pangu-weather [22, 43]), and the parallel architecture of AFNO and SWIN (introduced in Methods 4.2).

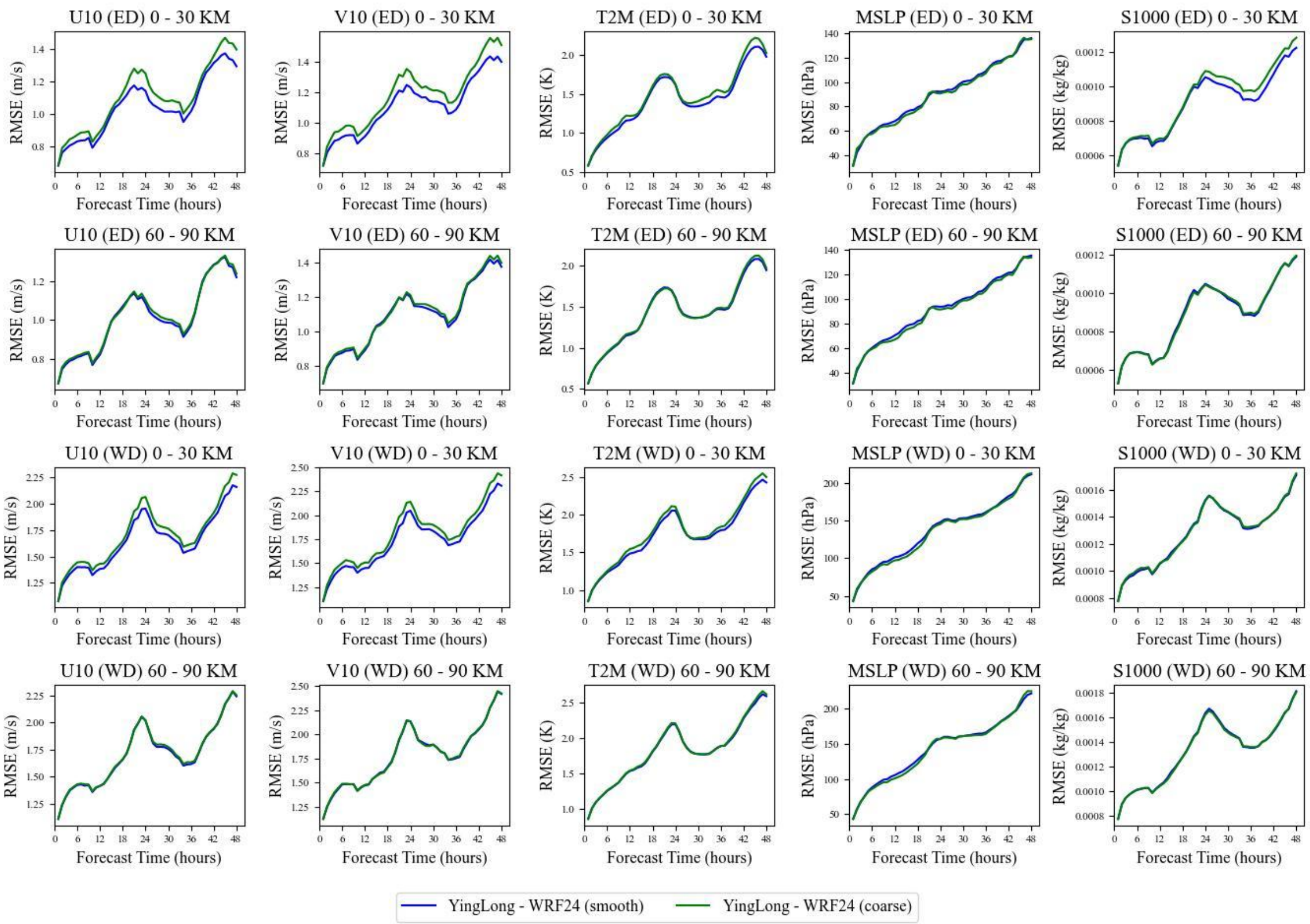

Figure 3. The impact of different lateral boundary strategies on the forecast effect at different ranges in the inner region. By selecting strip areas with distances of 0-30 km and 60-90 km from the inner region boundary, the performance of YingLong-WRF24 with boundary smoothing (blue lines) and direct replacement (green lines) are compared.

Intuitively, the parallel architecture of AFNO and SWIN can effectively extract multi-scale features of meteorological variables. Specifically, AFNO can extract the global features of meteorological variables, while SWIN can extract local features. In the parallel architecture, the $C$ channels of the neural network's hidden tensor are divided, with the first $\alpha * C$ channels subjected to the SWIN transformer and the remaining $(1-\alpha) * C$ channels subjected to AFNO. The two calculation results are then concatenated together (see Method 4.2 for details). This parallel architecture allows for the integration of global and local meteorological variable features, resulting in more accurate predictions. The hyperparameter $\alpha$ is estimated to be 0.25, as determined by the ablation test documented in Method 4.3.

### 3.2 Lateral boundary condition

YingLong can perform rolling forecasts for the entire area, including the internal area and the lateral boundary region, but only the forecast results in the internal area are the focus of this paper. As the forecast time increases, the analysis variables in the inner area are increasingly affected by the initial values outside the inner area. However, if the information of the variables outside the inner area is not included in the rolling forecast, it means that YingLong only uses the variables inside the domain as the initial field. Consequently, the prediction cannot account for the inner area being affected by the initial conditions outside of the domain in the next step. This could cause the difference between the YingLong forecast and the benchmark to become larger with increasing iterations of the rolling forecasts, as demonstrated in Fig. 4 (black lines). To solve this problem, a lateral boundary region is established outside the inner area. This region contains meteorological information from outside the inner area within a forecast time step $\Delta t$ of the rolling forecast, so it can affect the next forecast at time $t + \Delta t$ of meteorological variables in the inner area. The LBC can be provided by coarse-resolution global weather forecast models and can be transferred to the inner area through YingLong's forecast on the lateral boundary region (see Methods 4.4).

To obtain high-quality forecast results, it is necessary to ensure that the initial state of the lateral boundary at each step of the rolling forecast closely matches the actual atmospheric state. Directly using the coarse-resolution forecast on the lateral boundaries (coarse LBC, see section 4.4) at each step of the rolling forecast may cause discontinuities at the junction between the lateral boundary region and the inner area, potentially affecting the forecast results. However, by employing the boundary smoothing strategy (see Methods 4.4, Eq. (3)), the state of meteorological variables between the lateral boundary region and the inner area can be made more coherent, which could be beneficial for the rolling forecast.

Fig. 3 shows the difference in RMSEs (taking the analysis data as the benchmark) when using or not using the boundary smoothing strategy, as well as how this difference changes with the spatial range. It can be seen from Fig. 3 that within the 0-30 km strip from the inner area boundary (the red dashed box in Fig. 1b), there is a significant difference between the boundary smoothing strategy and the direct replacement strategy. As the area under consideration gets closer to the regional center, such as a strip area 60-90 km from the inner area boundary, the difference between these two boundary strategies is reduced. Within a range greater than 120 km from the inner zone boundary, there is almost negligible difference between the two boundary strategies. Therefore, different boundary strategies have an impact on AI-LAM forecasts, with a smaller impact closer to the domain center. In addition, the significance test (see Sec. SI 1.2 and Table S2 in SI 2 of Supplementary Material) also verified this conclusion.

The width of the lateral boundary region is related to how much information from outside the inner area is introduced into the LAM at each step. A principle for selecting this width is that the analysis data in the inner area at time $t + \Delta t$ should be related to the analysis in the lateral boundary region at time $t$, but not related to the analysis outside the domain at time $t$. If the width of the lateral boundary region is too small, the forecast of the inner area at $t + \Delta t$ will lack the necessary information from the lateral boundary region at time $t$, thus affecting the forecast accuracy for the inner area. Conversely, if the width is too large, some meteorological information in the lateral boundary region that has little to do with the rolling forecast could be introduced, resulting in wasted computational resources. The top speed of level 3 hurricanes in North America is 207 km/h. Unless there is a level 4 or 5 hurricane, the variability outside of the domain has no influence on the inner area within one hour. To validate the 207 km width against other shorter widths, i.e., 24 km and 96 km, RMSEs of the YingLong-WRF24 forecast in the inner area on smooth LBC with lateral boundary widths of 24 km, 96 km, and 207 km are shown in Fig. 4. The lateral boundary region with a width of 207 km corresponds to the smallest RMSEs. On the other hand, the RMSEs for the lateral boundary region with a width of 96 km are fairly close to those with a width of 207 km, indicating that 207 km is a reasonable choice.

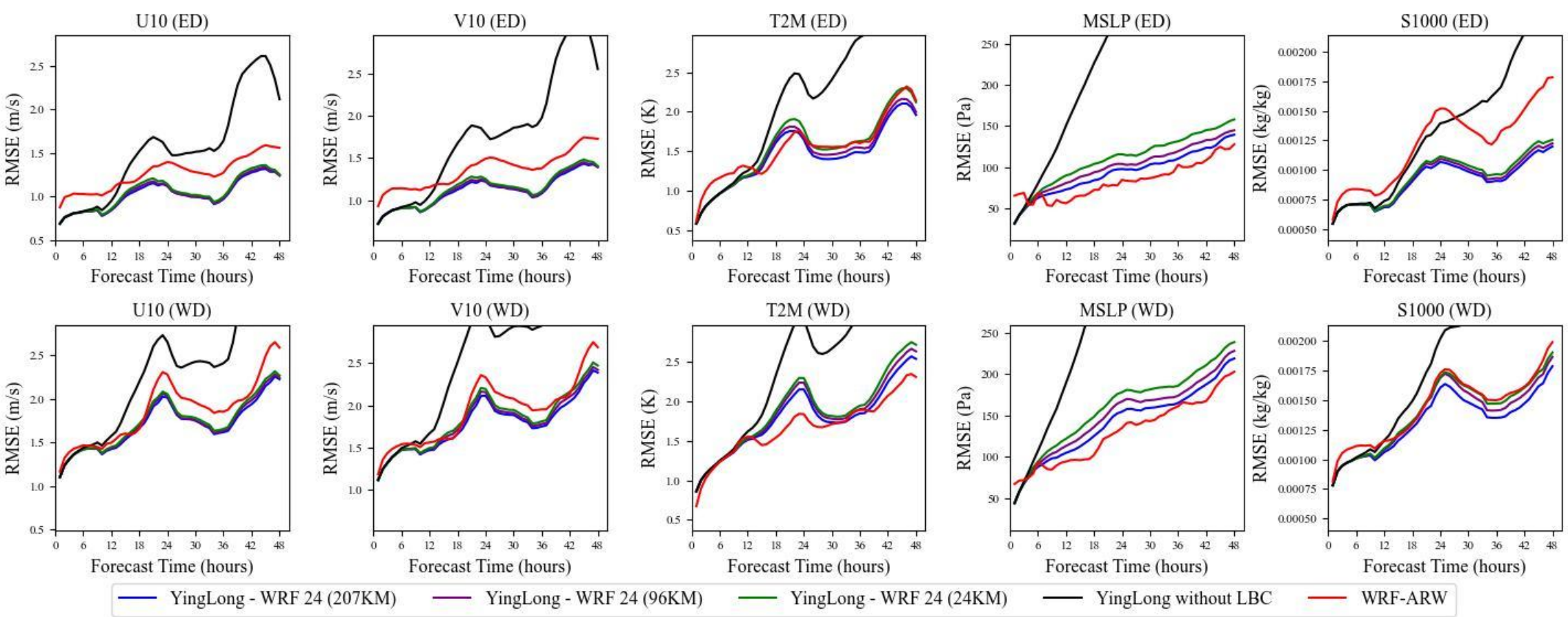

Figure 4. RMSEs (HRRR analysis as benchmark) of YingLong-WRF24 forecast in the inner area with no LBC (black lines), lateral boundary with widths 24 km (green lines), 96km (purple lines), 207 km (blue lines) and WRF-ARW (red line).

Figure 4 shows that the skills of YingLong-WRF24 are higher than those of WRF-ARW for predicting HRRR analysis U10 and V10, while this is not always the case for predicting HRRR analysis T2M and MSLP. This is similar to the skills for predicting station observations (Fig. 2). Moreover, the skills of YingLong-WRF24 are higher than those of WRF-ARW for predicting HRRR analysis S1000, especially in ED, suggesting that YingLong could be more skillful than WRF-ARW in forecasting surface specific humidity. The corresponding ACC results can also reach the same conclusion (see Fig. S5 in Supplementary Materials).

### 3.3 Effect of data homogenies

YingLong is constructed on two domains shown in Fig. 1(a), both with $440 \times 408$ grid points in Lambert projection. The training dataset spans the period from 2015 to 2021 and includes HRRR outputs from four different versions of the system (v1 - v4). The differences among these versions can be significant [44, 45], including changes in physics schemes, data assimilation methods, and land-surface data. The HRRR datasets for 2022 and 2023, used for testing and validation respectively, are in version v4 only.

To verify the impact of version inhomogeneity, two experiments were performed. The first experiment tested the sensitivity of YingLong's parameters to version inhomogeneity. The YingLong model was further fine-tuned on HRRR v4 data from 2021 to 2023 and tested on 2024 HRRR data. The RMSEs of the fine-tuned YingLong were very close to those of the original YingLong, indicating that the parameters of YingLong may not be very sensitive to version inhomogeneity in this study. The second experiment tested the reliability of using only HRRR v4 data for training. YingLong was trained on HRRR v4 data from 2021 to 2023 and tested on 2024 data. The RMSE of the test result was generally larger than that of YingLong trained using 7-year HRRR v1-v4 data (see Fig. S2 in SI 4 of Supplementary Material). This suggests that data length is more important than data homogeneity in this study. Therefore, training YingLong using 7-year HRRR v1-v4 data is a reasonable choice. The HRRR analysis data is hourly, resulting in 7*365*24 records used for training YingLong, which is roughly comparable to the ERA5 records (40*365*4) used for training global AI models.

### 3.4 Extreme events

Probability of Detection (POD, Eq. (S4) in SI 5 of Supplementary Material), False Alarm Ratio (FAR, Eq. (S5) of Supplementary Material), and Symmetric Extremal Dependence Index (SEDI, Eq. (S6) of Supplementary Material) are three metrics commonly used to evaluate the predictability of extreme events. A larger POD (or smaller FAR) indicates better predictability of extreme events. SEDI is an index that balances POD and the probability of false detection, with a higher SEDI value indicating better overall performance. In this study, wind speed at 10 meters above the surface ($\sqrt{\mathrm{U10}^2 + \mathrm{V10}^2}$) is selected for evaluating the predictability of extreme events for both YingLong and NWP. For WD, the threshold is set at 17.2 m/s (approximately the Beaufort wind gale scale [46, 47]). Since gale-scale wind is rare for the ED, it is difficult to estimate the three metrics accurately. Therefore, the threshold is reduced to 10.8 m/s (about a strong breeze). The POD, FAR, and SEDI of the wind over the two domains are shown in Fig. S3 in SI 5 of Supplementary Material. It shows that both POD and FAR for the YingLong forecast are lower than those of NWP. Additionally, the SEDIs of the YingLong forecast in both domains are larger than those of WRF-ARW overall, suggesting that the YingLong performs better than WRF-ARW in detecting extreme events. These results are consistent with those of global AI-based models [48].

## 4. Methods

### 4.1 Datasets

The hourly analysis data of the HRRR dataset [44, 45, 49] at 3-km resolution, covering the continental US and Alaska, is used for training and evaluating the YingLong model. The HRRR dataset also provides NWP at 3-km resolution, generated through the WRF-ARW. The hourly NWP starts from 00UTC and continues for 48 hours.

To test the forecast skill of YingLong-Pangu, the *in-situ* observation data from HadISD [39] in 2022 are selected as benchmarks for calculating RMSEs and ACC. The *in-situ* observation data includes hourly T2M, U10, V10, and MSLP. This study uses a total of 205 observational sites, with 78 sites in WD and 127 sites in ED.

### 4.2 Architecture

The YingLong consists of three components shown in Fig. 1(c): embedding layer, spatial mixing layers, and a linear decoder. The input data for the embedding layer are the 24 variables listed in Table I and one forcing elevation. They form a tensor with dimensions of $440 \times 408 \times 25$. The embedding layer consists of patch embedding, position embedding and time embedding, to integrate both the spatial and temporal information into the latent tensor. Patch embedding partitions the input tensor into 2805 patches, each patch with a size $8 \times 8 \times 25$. Then, through a convolution layer, each patch is encoded into a $C$ dimensional vector, resulting in the entire input variables being encoded as a tensor of size $2805 \times C$. Position embedding consists of 2805 learnable parameter vectors associated with each of those 2805 relative positions, enabling the model to learn the encoding for each relative. position in the region. The dimension of the position embedding vectors is also set to C. Time embedding encodes the specific time information of input data, including year, month, day, and hour, in a vector of dimensions C. Summing up the output vectors from the patch embedding, the position embedding, and the time embedding, yields the output of the embedding layer of size $2805 \times C$. In this study, $C = 768$.

The output of the embedding layer is reshaped into a $55 \times 51 \times C$ tensor and delivered to the spatial mixing layer. The tensor is further split by a ratio α along the channel dimension. The tensor of size $55 \times 51 \times (\alpha \cdot C)$ is delivered to the "Local" branch, and the remaining tensor of size $55 \times 51 \times ((1-\alpha) \cdot C)$ is sent to the "Global" branch. The "Local" and "Global" branches operate independently and in parallel. The outputs from the two branches are concatenated along the channel dimension, and a new tensor with a dimension of is $55 \times 51 \times C$ obtained. The "Local" branch contains Window Multi-head Self-Attention (W-MSA) to capture the features within the window, and Shift Window Multi-head Self-Attention (SW-MSA) blocks in the Swin Transformer [37] to find the relationships among different windows by a shifting operation. For W-MSA, padding is performed so that the $55 \times 51 \times (\alpha \cdot C)$ tensor is transferred to $56 \times 56 \times (\alpha \cdot C)$ tensor. Then it can be divided by the 8 × 8 window to get 7 × 7 patches. For SW-MSA, the window is shifted by three patches at each time. After a few alternating steps of W-MSA and SW-MSA, the padding data is removed and a tensor of size $55 \times 51 \times (\alpha \cdot C)$ is returned. The "Global" branch mainly utilizes the Adaptive Fourier Neural Operator (AFNO) block [36]. A 2D fast Fourier transform is applied to the area of 55 × 51 along each $(1-\alpha) \cdot C$ channel. In the Fourier frequency domain, feature mixing is carried out using a Multilayer Perceptions (MLP) consisting of two linear layers. Information from the frequency domain is transferred to the spatial domain by inverse fast Fourier transform. The outputs of the two branches are concatenated along the channel dimension, and a new tensor with a dimension of $55 \times 51 \times C$ can be given.

An MLP consisting of two linear layers is applied to map the channel dimension from $C$ to (8 × 8 × 24) for Linear Decoder. The tensor of size 55×51×(8×8×24) is then reshaped back to 440×408×24 as the final output of the YingLong as the forecast result at $t + \Delta t$ with $\Delta t = 1$h.

### 4.3 Training

The input for YingLong pretraining, denoted as $X_t$, is the HRRR analysis in 2015-2021 and the elevation. The output YingLong$(X_t)$ is denoted by $\hat{X}_{t+\Delta t}$, the analysis data $X_{t+\Delta t}$ at the corresponding time is used as labels for supervised learning. The parameters of YingLong are updated by minimizing the loss function

$$\mathcal{L}_1 = \frac{1}{|D_{batch}|} \sum_{t \in D_{batch}} \frac{|| \hat{X}_{t+\Delta t} - X_{t+\Delta t} ||_2}{||X_{t+\Delta t}||_2}, \tag{1}$$

where the second norm of the vector $X = (x_1, x_2, \ldots, x_n)$ is expressed as $\|X\|_2 = \sqrt{x_1^2 + x_2^2 + \cdots + x_n^2}$, $t \in D_{batch}$ is the initial forecast time in a minibatch of forecasts in the training set. For the definition of minibatch, please refer to Chapter 11 of [50]. To achieve better predictability at longer forecast time, the pretrained parameters are fine-tuned by minimizing the loss function

$$\mathcal{L}_2 = \frac{1}{|D_{batch}|} \sum_{t \in D_{batch}} \sum_{i=1}^{T} \frac{|| \hat{X}_{t+i\Delta t} - X_{t+i\Delta t} ||_2}{||X_{t+i\Delta t}||_2}, \tag{2}$$

where $\hat{X}_{t+(i+1)\Delta t} = YingLong(\hat{X}_{t+i\Delta t})$ for $i \geq 1$. In this study, $T = 6$. The larger time step $i$, the more error is induced to $\hat{X}_{t+i\Delta t}$ from outside of the domain. Based on the trial, $T > 6$ not only fails to improve YingLong's long-range forecast skill, but also requires more computational resources.

The Adam optimizer [50,51] is utilized to minimize the two loss functions. The model employs a cosine learning rate schedule for pre-training, starting at an initial learning rate of 0.005, iterated over 30 epochs. Following pre-training, the model is fine-tuned for 15 epochs using a cosine learning rate schedule with a lower learning rate of 0.0001. The YingLong model is trained on two A100 GPUs, taking approximately 7 days. The trained YingLong model takes about 0.5 seconds to generate a 48-hour hourly forecast result on a single A100 GPU.

In this study, the hyper parameter $\alpha$ is selected from 0 (AFNO), 0.25, 0.75, and 1 (SWIN) by conducting ablative tests. For each $\alpha$, the parameters of YingLong are estimated using HRRR analysis in 2015-2021; then the RMSEs of the YingLong-WRF24 forecast with HRRR as benchmark in the training period, the validation period 2023 and testing period 2022 are shown in Fig. S1a-c in SI 2 respectively. In ED RMSEs for $\alpha = 0.25$ are overall smallest for all the periods. This is also the case in WD, but with less significance to the case of $\alpha = 0$ (SWIN). The number of parameters of the AI-based models are listed in Table 2. It shows that the number of the parameters of YingLong (60.5M) is less than that of SWIN (85.0 M), so YingLong is trained more economically than SWIN. For all of these reasons, $\alpha = 0.25$ (YingLong) is selected by the ablative test.

Table 2

| Model | Number of parameters |
| --- | --- |
| YingLong ($\alpha = 0.25$) | 60.5M |
| SWIN ($\alpha = 1$) | 85.0M |
| AFNO ($\alpha$=0) | 60.3M |

**4.4 Lateral boundary condition**

To run a LAM requires an LBC $\hat{Y}_t$, which is normally obtained from AI-based global forecast or coarse resolution NWP, given by Pangu-weather or coarsen HRRR to 24 km depending on the purpose for forecast or research application. Two LBC schemes are tested, coarse LBC and smooth LBC. For the coarse LBC scheme, a forecast at fine resolution grid *n* within the lateral boundary region ($\hat{X}_t(n)$) is altered as $\hat{Y}_t(m)$ where *m* is each coarse resolution grid in the lateral boundary region that is nearest to the fine resolution grid *n*. For the smooth LBC scheme, it is further smoothed by the YingLong forecast $\hat{X}_t(n)$ based on the inverse distance weighting approach on the lateral boundary. Specifically, for any point $n$ in the lateral boundary, there is a YingLong forecast result $\hat{X}_t(n)$, and the coarse-resolution NWP or AI-based global forecast result $\hat{Y}_t(m)$ at the point $m$ closest to $n$ in the coarse-resolution grid.

When the distance from point $n$ to the corresponding inner area, denoted as $d(n)$, is less than the width of the lateral boundary region, which is 207 km in this paper, the result of boundary smoothing $\mathcal{X}_t(n)$ can be expressed as

$$\mathcal{X}_t(n) = (1 - d(n)/207)\hat{X}_t(n) + d(n)/207\hat{Y}_t(m). \quad (3)$$

When $d(n)$ is greater than the width of the lateral boundary (this is the case in the four corners of lateral boundary area), i.e., 207 km, $\mathcal{X}_t(n) = \hat{Y}_t(m)$. The closer the fine resolution grid $n$ to the inner area, the more weight of YingLong forecast at the grid.

### 4.5 Rolling forecast

Take ED as an example to introduce the process of YingLong rolling forecast. Let $X_t(\mathrm{ED})$, $X_t(\mathrm{ED_{in}})$, $X_t(\mathrm{ED_{out}})$ and $X_t(\mathrm{ED_{LB}})$ represent states in entire, inner, outside and lateral boundary regions of ED at time $t$ respectively. Firstly, YingLong maps an initial state $X_t(\mathrm{ED})$ to the forecast state

$$\hat{X}_{t+\Delta t}(\mathrm{ED}) = YingLong\big(X_t(\mathrm{ED})\big).$$

Then $\hat{X}_{t+\Delta t}(\mathrm{ED_{LB}})$ and a coarse resolution LBC $\hat{Y}_{t+\Delta t}(\mathrm{ED_{LB}})$ are smoothed by Eq (3) to form the initial state $X_{t+\Delta t}(\mathrm{ED})$ for the next time step (see Fig 1(a)).

## Data availability

The training and testing data for the YingLong model, we download a part of the HRRR dataset from https://rapidrefresh.noaa.gov/hrrr/, and this website can also provide the forecasting results of various variables with WRF-ARW. Pangu-weather forecast data can be downloaded from https://weatherbench2.readthedocs.io/en/latest/data-guide.html#forecast-datasets. The weather station dataset HadISD can be found at https://www.metoffice.gov.uk/hadobs/hadisd/.

## Code availability

The code architecture of YingLong is developed on PaddlePaddle, a Python-based framework for deep learning, available at https://github.com/PaddlePaddle/Paddle. During building our architecture, we utilize part of the Swin transformer, see https://github.com/microsoft/Swin-Transformer, the AFNO block is also involved, which can be found in https://github.com/NVlabs/AFNO-transformer. The trained YingLong models and some details are released in a GitHub repository: https://github.com/PaddlePaddle/PaddleScience/tree/develop/examples/yinglong

## Acknowledgment

This work was supported by Major Program of National Natural Science Foundation of China (NSFC) Nos.12292980, 12292984, NSFC 12405038, NSFC12031016, NSFC12201024, NSFC12426529; National Key R&D Program of China Nos.2023YFA1009000, 2023YFA1009004, 2020YFA0712203, 2020YFA0712201; Beijing Natural Science Foundation BNSF-Z210003; was funded by the Department of Science, Technology and Information of the Ministry of Education, No:8091B042240, and was also Sponsored by CCF-Baidu Open Fund CCF-BAIDU 202317.

## Author contributions

P.X., J.Y., and J.Z. designed the project. P.X. and J.Z. managed and supervised the project. P.X., T.G., and Y.W. carried out model design, development, training, and testing. X.Z. guided the model's meteorological experiments. P.X., and X.Z. wrote the manuscript. X.Z., S.L., and Z.W., helped revise the manuscript and polish the language. Z.Z., X.H., and X.C. deployed the training environment required by the model.

## Competing interests

The authors declare no competing interests.

**Supplemental Material**

## SI 1 Statistical test

### S1 1.1 Evaluation metrics

The root mean squared error (RMSE) and anomaly correlation coefficient (ACC) are two important evaluation metrics widely used in various models, such as FourCastNet, Pangu-weather, etc. They are defined as

$$\mathrm{RMSE} = \frac{1}{N}\sum_{k=1}^{N}\sqrt{\frac{1}{H\times W}\sum_{x=1}^{H}\sum_{y=1}^{W}\left(\hat{X}_{x,y,k}-X_{x,y,k}\right)^{2}}, \tag{S1}$$

$$\mathrm{ACC} = \frac{\sum_{k=1}^{N}\sum_{x=1}^{H}\sum_{y=1}^{W}(\hat{X}_{x,y,k}-\bar{X}_{x,y})(X_{x,y,k}-\bar{X}_{x,y})}{\sqrt{\sum_{k=1}^{N}\sum_{x=1}^{H}\sum_{y=1}^{W}\left(\hat{X}_{x,y,k}-\bar{X}_{x,y}\right)^{2}\sum_{k=1}^{N}\sum_{x=1}^{H}\sum_{y=1}^{W}\left(X_{x,y,k}-\bar{X}_{x,y}\right)^{2}}} \tag{S2}$$

where $\bar{X}_{x,y} = \frac{1}{M}\sum_{k'=1}^{M}\hat{X}_{x,y,k'}$ is the climatology averaging on each variable from 2015 to 2022, and $M$ is the number of samples in the whole dataset, $N$ represents the number of whole test samples, and $H$ (=302 pixels) and $W$ (=270 pixels) are the height and width of each variable in the inner domain, respectively. The higher ACC and lower RMSE stand for the better forecasting ability of the corresponding model.

### SI 1.2 significance tests for difference in means

The significance test method used in this article is similar to that of GraphCast (see Sec. 5.4.1 in [47]), see details in [48]. For each forecast time $t$ and variable $v$, we test for a difference in means between each RMSEs for smooth LBC, coarse LBC and NWP initialization-time $i$. Take the significance test of NWP (i.e. WRF-ARW) and AI (i.e. YingLong) forecast models as an example. Define

$$\mathbf{H_0}\text{:}\ \mu_{nwp} = \mu_{AI}\ \textbf{vs}\quad \mathbf{H_1}\text{:}\ \mu_{nwp} \neq \mu_{AI},$$

where $\mu$ is the mean of RMSEs.

For two sets of samples RMSE($i$) ($i = 1,2,\ldots,N$), since the results of NWP and AI for the same moment $i$ are paired, we use a paired $t$-test. The designed statistical measure is as follows,

$$t = \frac{\bar{d}}{s_d/\sqrt{n}}, \tag{S3}$$

Calculate the average of all the differences $\bar{d} = \frac{1}{n}\sum_{i=1}^{N} di$, $di$ represents the difference between the $i$-th pair of samples $di = RMSE_{nwp}(i) - RMSE_{AI}(i)$, $s_d = \sqrt{\frac{\sum_{i-1}^{n}(d_i - \bar{d})^2}{n-1}}$.

As shown by Fisher (1998), and more generally by, for example, Wilks (2006), in the $t$-test of the paired differences in forecast scores, it is possible to account for the effect of temporal autocorrelation using autoregression (AR) models. Temporal autocorrelation increases the variance of the sampling distribution of the mean relative to the variance of the sample, which can be represented with an inflation factor $k$ and formulated as

$$\mathrm{Var}\,[\bar{d}] = \frac{k^2 s^2}{n} \quad \text{(S4)}$$

and the sample mean $d$, sample standard deviation $s$ and sample size $n$. The $t$-test computes the ratio of the sample mean to its estimated sampling error, so the inflation factor k describes the increase in that expected sampling error.

These correspond to reduced effective sample sizes ($ess = n/k^2$).

## SI 2 Ablation Experiments

In this part, these experiments remain consistent in other parameter settings. Figure S1 presents the ablation experiments regarding the parallel parameter ratio of AFNO and Swin. Ratio $\alpha$ value of zero indicates the exclusive use of the AFNO, while ratio value of 1 indicates the exclusive use of the Swin Transformer. It can be observed from Fig. S1 that when the parameter is set to 0.25, the YingLong model achieves the best overall performance in the validation set (2023) and the test set (2022), across all surface variables both in ED and WD. It should be noted that the corresponding analysis data of HRRR is used as the benchmark for calculating RMSE and ACC.

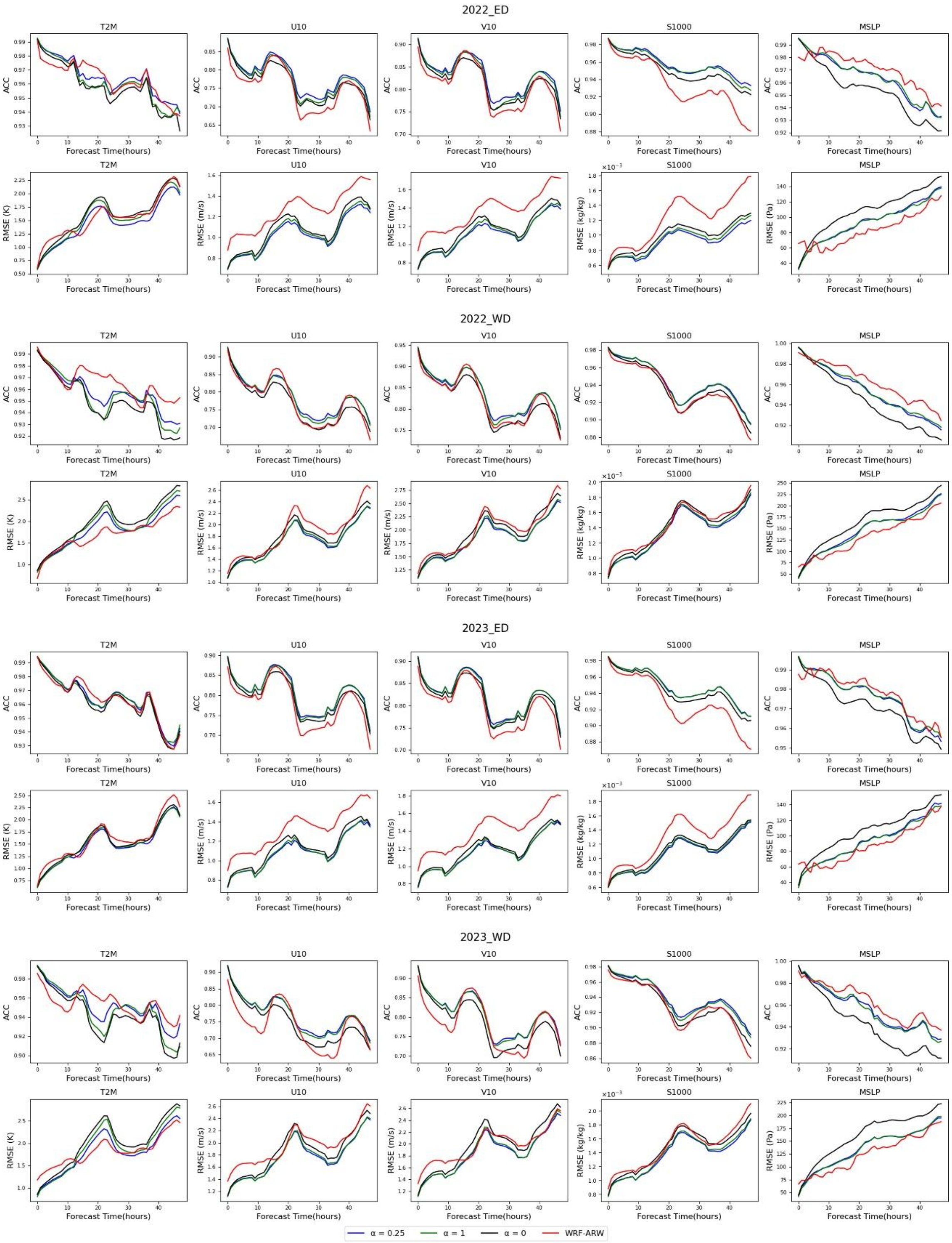

Figure S1: Ablation experiments for different choices of $\alpha$. Select HRRR analysis data as benchmark for calculating RMSE and ACC. For the case of $\alpha = 0.25$, the corresponding model is YingLong used in this paper. For the cases of $\alpha = 0$ and $\alpha = 1$, they correspond to the AFNO and SWIN transformer architectures, respectively.

**Table S1:** Significant test of difference of RMSEs with ablation experiments.

Validation for 2023 in ED:

| YingLong - SWIN (2023) | | | | | | | | | |
|---|---|---|---|---|---|---|---|---|---|
| forecast time | 12h | | | 24h | | | 48h | | |
| variable | $t$ statistic | $p$ value | ess | $t$ statistic | $p$ value | ess | $t$ statistic | $p$ value | ess |
| U10 | -1.008 | 3.143E-01 | 360.553 | -1.932 | 5.42E-02 | 357.571 | -1.496 | 1.356E-01 | 358.440 |
| V10 | -0.067 | 9.466E-01 | 355.936 | -1.279 | 2.02E-01 | 355.164 | -0.176 | 8.607E-01 | 361.839 |
| T2M | -2.693 | 7.403E-03 | 342.367 | -2.933 | 3.57E-03 | 350.210 | -1.107 | 2.690E-01 | 347.532 |
| S1000 | -5.628 | 3.640E-08 | 319.779 | -2.488 | 1.33E-02 | 307.089 | -3.609 | 3.508E-04 | 311.488 |
| YingLong - AFNO (2023) | | | | | | | | | |
| forecast time | 12h | | | 24h | | | 48h | | |
| variable | $t$ statistic | $p$ value | ess | $t$ statistic | $p$ value | ess | $t$ statistic | $p$ value | ess |
| U10 | -2.568 | 1.062E-02 | 360.702 | -3.404 | 7.38E-04 | 356.351 | -2.096 | 3.679E-02 | 360.150 |
| V10 | -1.459 | 1.453E-01 | 359.112 | -2.381 | 1.78E-02 | 355.428 | -0.524 | 6.009E-01 | 360.865 |
| T2M | -4.095 | 5.200E-05 | 352.911 | -4.037 | 6.61E-05 | 347.921 | -2.819 | 5.078E-03 | 329.259 |
| S1000 | -8.368 | 1.270E-15 | 326.037 | -5.453 | 9.16E-08 | 295.970 | -5.780 | 1.610E-08 | 302.639 |

Test for 2022 in ED:

| YingLong - SWIN (2022) | | | | | | | | | |
|---|---|---|---|---|---|---|---|---|---|
| forecast time | 12h | | | 24h | | | 48h | | |
| variable | $t$ statistic | $p$ value | ess | $t$ statistic | $p$ value | ess | $t$ statistic | $p$ value | ess |
| U10 | -3.318 | 9.995E-04 | 364.135 | -5.776 | 1.64E-08 | 364.974 | -4.503 | 9.050E-06 | 364.635 |
| V10 | -3.315 | 1.007E-03 | 364.737 | -7.120 | 5.79E-12 | 364.748 | -4.868 | 1.690E-06 | 362.481 |
| T2M | -5.481 | 7.900E-08 | 323.823 | -4.894 | 1.49E-06 | 329.885 | -2.702 | 7.213E-03 | 347.666 |
| S1000 | -5.551 | 5.470E-08 | 351.840 | -3.419 | 7.00E-04 | 340.101 | -6.248 | 1.160E-09 | 352.304 |
| YingLong - AFNO (2022) | | | | | | | | | |
| forecast time | 12h | | | 24h | | | 48h | | |
| variable | $t$ statistic | $p$ value | ess | $t$ statistic | $p$ value | ess | $t$ statistic | $p$ value | ess |
| U10 | -11.191 | 3.490E-25 | 362.330 | -8.055 | 1.14E-14 | 363.747 | -4.877 | 1.610E-06 | 362.112 |
| V10 | -9.549 | 1.980E-19 | 361.709 | -9.006 | 1.22E-17 | 360.549 | -4.063 | 5.930E-05 | 364.062 |
| T2M | -11.110 | 6.840E-25 | 362.404 | -8.680 | 1.34E-16 | 357.842 | -6.566 | 1.790E-10 | 354.209 |
| S1000 | -13.194 | 9.300E-33 | 364.146 | -9.570 | 1.68E-19 | 363.086 | -8.487 | 5.440E-16 | 359.286 |

Validation for 2023 in WD:

| YingLong - SWIN (2023) | | | | | | | | | |
|---|---|---|---|---|---|---|---|---|---|
| forecast time | 12h | | | 24h | | | 48h | | |
| variable | $t$ statistic | $p$ value | ess | $t$ statistic | $p$ value | ess | $t$ statistic | $p$ value | ess |
| U10 | -1.687 | 9.25E-02 | 332.708 | -3.357 | 8.72E-04 | 330.727 | -2.346 | 1.95E-02 | 337.781 |
| V10 | -0.295 | 7.68E-01 | 332.099 | -1.226 | 2.21E-01 | 361.969 | 0.079 | 9.37E-01 | 362.130 |
| T2M | 4.042 | 6.48E-05 | 341.833 | -6.018 | 4.30E-09 | 313.907 | -2.793 | 5.50E-03 | 334.563 |
| S1000 | -2.075 | 3.87E-02 | 300.287 | -1.231 | 2.19E-01 | 314.619 | -1.451 | 1.48E-01 | 323.639 |
| YingLong - AFNO (2023) | | | | | | | | | |
| forecast time | 12h | | | 24h | | | 48h | | |
| variable | $t$ statistic | $p$ value | ess | $t$ statistic | $p$ value | ess | $t$ statistic | $p$ value | ess |
| U10 | -3.850 | 1.39E-04 | 332.906 | -5.883 | 9.12E-09 | 328.518 | -3.982 | 8.25E-05 | 335.219 |
| V10 | -3.106 | 2.04E-03 | 340.140 | -3.759 | 1.99E-04 | 363.000 | -1.511 | 1.32E-01 | 363.384 |
| T2M | -1.939 | 5.33E-02 | 336.631 | -6.790 | 4.56E-11 | 284.971 | -3.194 | 1.52E-03 | 309.512 |
| S1000 | -5.423 | 1.07E-07 | 295.138 | -3.421 | 6.94E-04 | 307.704 | -3.267 | 1.19E-03 | 321.428 |

Test for 2022 in WD:

| YingLong - SWIN (2022) | | | | | | | | | |
|---|---|---|---|---|---|---|---|---|---|
| forecast time | 12h | | | 24h | | | 48h | | |
| variable | $t$ statistic | $p$ value | ess | $t$ statistic | $p$ value | ess | $t$ statistic | $p$ value | ess |
| U10 | -8.889 | 2.90E-17 | 364.775 | -6.842 | 3.31E-11 | 361.329 | -6.182 | 1.70E-09 | 359.719 |
| V10 | -8.851 | 3.84E-17 | 362.835 | -7.922 | 2.88E-14 | 364.970 | -7.174 | 4.10E-12 | 364.980 |
| T2M | -7.223 | 2.99E-12 | 349.768 | -9.033 | 9.95E-18 | 334.876 | -7.236 | 2.76E-12 | 342.604 |
| S1000 | -8.399 | 1.02E-15 | 364.465 | -6.217 | 1.39E-09 | 358.832 | -5.100 | 5.48E-07 | 357.612 |
| YingLong - AFNO (2022) | | | | | | | | | |
| forecast time | 12h | | | 24h | | | 48h | | |
| variable | $t$ statistic | $p$ value | ess | $t$ statistic | $p$ value | ess | $t$ statistic | $p$ value | ess |
| U10 | -8.889 | 2.90E-17 | 364.775 | -6.842 | 3.31E-11 | 361.329 | -6.182 | 1.70E-09 | 359.719 |
| V10 | -8.851 | 3.84E-17 | 362.835 | -7.922 | 2.88E-14 | 364.970 | -7.174 | 4.10E-12 | 364.980 |
| T2M | -7.223 | 2.99E-12 | 349.768 | -9.033 | 9.95E-18 | 334.876 | -7.236 | 2.76E-12 | 342.604 |
| S1000 | -8.399 | 1.02E-15 | 364.465 | -6.217 | 1.39E-09 | 358.832 | -5.100 | 5.48E-07 | 357.612 |

According to the statistics of the number of parameters of the models corresponding to different $\alpha$, it can be found that when $\alpha = 0.25$ (corresponds to the YingLong model in this paper), the number of parameters is far less than that of SWIN and slightly more than that of AFNO. The prediction of the model with $\alpha = 0.25$ is the best in the set of [0,0.25,1].

**SI 3 Lateral boundary condition**

According to the significance test (see Table S2), it can be generally found that there are significant differences between different boundary strategies at a distance of 0-30km from the inner area boundary, and the boundary smoothing strategy has better prediction results than directly replacing the boundary. However, within 60-90km to the boundary of inner domain, the significance of the difference between different boundary strategies for the forecast model decreases.

Table S2: Significant test of difference of RMSEs of Figure 3.

**ED:**

| YingLong Smooth - Coarse (0-30 km) | | | | | | | | | |
|---|---|---|---|---|---|---|---|---|---|
| forecast time | 12h | | | 24h | | | 48h | | |
| variable | $t$ statistic | $p$ value | ess | $t$ statistic | $p$ value | ess | $t$ statistic | $p$ value | ess |
| U10 | -10.454 | 1.530E-22 | 361.325 | -17.962 | 4.19E-52 | 334.934 | -17.506 | 3.250E-50 | 360.062 |
| V10 | -14.204 | 9.640E-37 | 363.036 | -17.379 | 1.08E-49 | 358.836 | -16.543 | 3.080E-46 | 364.529 |
| T2M | -10.564 | 6.250E-23 | 356.065 | -0.515 | 6.07E-01 | 362.204 | -3.924 | 1.043E-04 | 355.980 |
| S1000 | -5.496 | 7.300E-08 | 364.993 | -7.352 | 1.30E-12 | 363.033 | -9.679 | 7.240E-20 | 353.990 |
| YingLong Smooth - Coarse (60-90 km) | | | | | | | | | |
| forecast time | 12h | | | 24h | | | 48h | | |
| variable | $t$ statistic | $p$ value | ess | $t$ statistic | $p$ value | ess | $t$ statistic | $p$ value | ess |
| U10 | -3.872 | 1.281E-04 | 362.364 | -4.029 | 6.83E-05 | 361.308 | -4.178 | 3.690E-05 | 364.360 |
| V10 | -3.960 | 9.010E-05 | 364.393 | -2.733 | 6.59E-03 | 364.497 | -4.785 | 2.480E-06 | 356.535 |
| T2M | -5.081 | 6.030E-07 | 363.989 | 1.309 | 1.91E-01 | 358.346 | -2.322 | 2.076E-02 | 357.073 |
| S1000 | 0.946 | 3.447E-01 | 362.137 | 1.400 | 1.62E-01 | 364.992 | -1.087 | 2.778E-01 | 361.959 |

**WD:**

| YingLong Smooth - Coarse (0-30 km) | | | | | | | | | |
|---|---|---|---|---|---|---|---|---|---|
| forecast time | 12h | | | 24h | | | 48h | | |
| variable | $t$ statistic | $p$ value | ess | $t$ statistic | $p$ value | ess | $t$ statistic | $p$ value | ess |
| U10 | -12.814 | 2.78E-31 | 364.998 | -10.009 | 5.43E-21 | 360.699 | -11.441 | 4.22E-26 | 362.168 |
| V10 | -10.826 | 7.27E-24 | 364.954 | -8.020 | 1.47E-14 | 361.135 | -9.560 | 1.82E-19 | 364.597 |
| T2M | -11.344 | 9.56E-26 | 363.324 | -3.477 | 5.69E-04 | 362.432 | -4.671 | 4.23E-06 | 350.269 |
| S1000 | -1.362 | 1.74E-01 | 357.669 | 0.625 | 5.33E-01 | 361.740 | -1.583 | 1.14E-01 | 362.251 |
| YingLong Smooth - Coarse (60-90 km) | | | | | | | | | |
| forecast time | 12h | | | 24h | | | 48h | | |
| variable | $t$ statistic | $p$ value | ess | $t$ statistic | $p$ value | ess | $t$ statistic | $p$ value | ess |
| U10 | -0.166 | 8.68E-01 | 364.901 | 0.176 | 8.60E-01 | 363.800 | -1.313 | 1.90E-01 | 362.933 |
| V10 | -1.278 | 2.02E-01 | 364.817 | 0.498 | 6.19E-01 | 364.929 | -0.562 | 5.75E-01 | 364.885 |
| T2M | -1.501 | 1.34E-01 | 364.417 | -0.905 | 3.66E-01 | 360.212 | -1.761 | 7.91E-02 | 353.793 |
| S1000 | 1.667 | 9.64E-02 | 361.024 | 1.835 | 6.73E-02 | 359.639 | 1.130 | 2.59E-01 | 357.200 |

## SI 4 Inhomogeneity of training data

Since HRRR v4 data is limited (2021 to 2024), in order to make full use of HRRR v4 data for training, this paper uses two methods. The first method is to use the trained YingLong model (the training set for YingLong is HRRR data from 2015 to 2021) to further fine-tune on HRRR from 2021 to 2023. The second method is to train the model from beginning on HRRR data from 2021 to 2023. The three models (YingLong model, model fine-tuned on HRRR v4 data, and model trained only on HRRR v4 data) are then tested on HRRR v4 data in 2024. According to the Fig. S2, the test results of fine-tuning on the v4 data are very close to YingLong, while the model trained only on the v4 data has larger RMSEs on T2M, U10, and V10. It should be noted that the corresponding analysis data of HRRR is used as the benchmark for calculating RMSE and ACC.

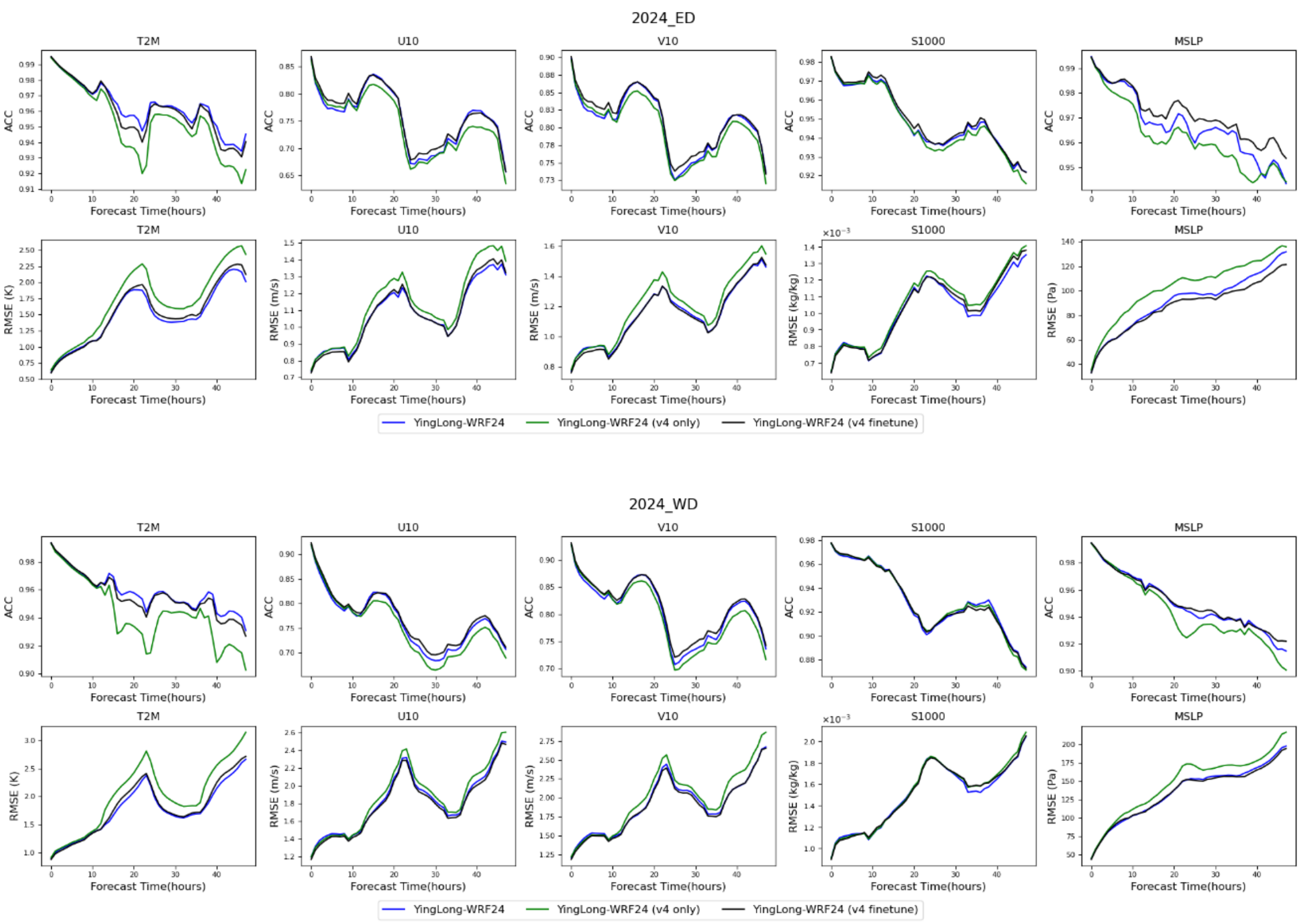

Figure S2. Impact of data version inhomogeneity. The RMSE and ACC are calculated based on the corresponding analysis data. By comparing YingLong on the entire training set (blue line), a model trained only on HRRR v4 (green line), and YingLong further fine-tuned on v4 (black line), it can be found that for the problem considered in this paper, the amount of training data is more important than the difference in versions.

## SI 5 Extreme events

### SI 5.1 Metric Indicators for Extreme Events

Before explaining the metrics of evaluating extreme events, we define the following four concepts: hit (model forecasts an extreme event, and an extreme event actually occurs), false alarm (model forecasts an extreme event, but no extreme event occurs in reality), miss (model forecasts no extreme event, but an extreme event actually occurs), and correct negatives (model forecasts no extreme event, and no extreme event occurs in reality). We use *a, b, c*, and *d* to denote the number of hits, false alarms, misses, and correct negatives in the forecasts, respectively.

Probability of detection (POD), also known as hit rate, is a critical metric used to assess the accuracy and reliability of forecasting extreme events. A higher POD indicates a stronger ability to forecast extreme events accurately, while a lower POD suggests more missed detection of actual extreme events, meaning the events were not detected when they actually occurred. The POD can be expressed as:

$$\mathrm{POD} = \frac{a}{a+c} \tag{S4}$$

its value ranges from 0 to 1, with higher values indicating a greater hit rate in forecasting extreme events.

The False Alarm Ratio (FAR) is a metric used to assess the rate of false alarms in weather forecasting for extreme events. FAR represents the proportion of false alarms (instances where the model predicts an extreme event that does not actually occur) out of the total number of alarms issued by the model. The FAR can be expressed as

$$\mathrm{FAR} = \frac{b}{a+b}, \tag{S5}$$

its value ranges from 0 to 1, with smaller values indicating a lower false alarm rate in forecasting extreme events.

The Symmetric Extremal Dependence Index (SEDI) is a metric that is used to evaluate the accuracy of the prediction of extreme events. SEDI takes into account both the hit rate and the false alarm rate, providing a comprehensive assessment of the prediction accuracy for extreme events. The SEDI can be expressed as:

$$\mathrm{SEDI} = \frac{\ln F - \ln H}{\ln F + \ln H}, \tag{S6}$$

where $F = b/(d+b)$, $H = a/(a+c)$, its values range from $-1$ to 1, with higher values closer to 1 indicating better forecast performance. In fact, it can be found that $H$ in (S4) is POD, and $F$ is false detection rate.

### SI 5.2 Results for Extreme Events

In the ED dominated by plains, extreme events are defined as wind speeds greater than 10.8 m/s, and in the WD rich in mountains, extreme events are defined as wind speeds > 17.2 m/s. Based on the corresponding analysis data, the POD, FAR and SEDI of YingLong and WRF-ARW for extreme event forecasts are calculated respectively. The results are shown in Fig. S3.

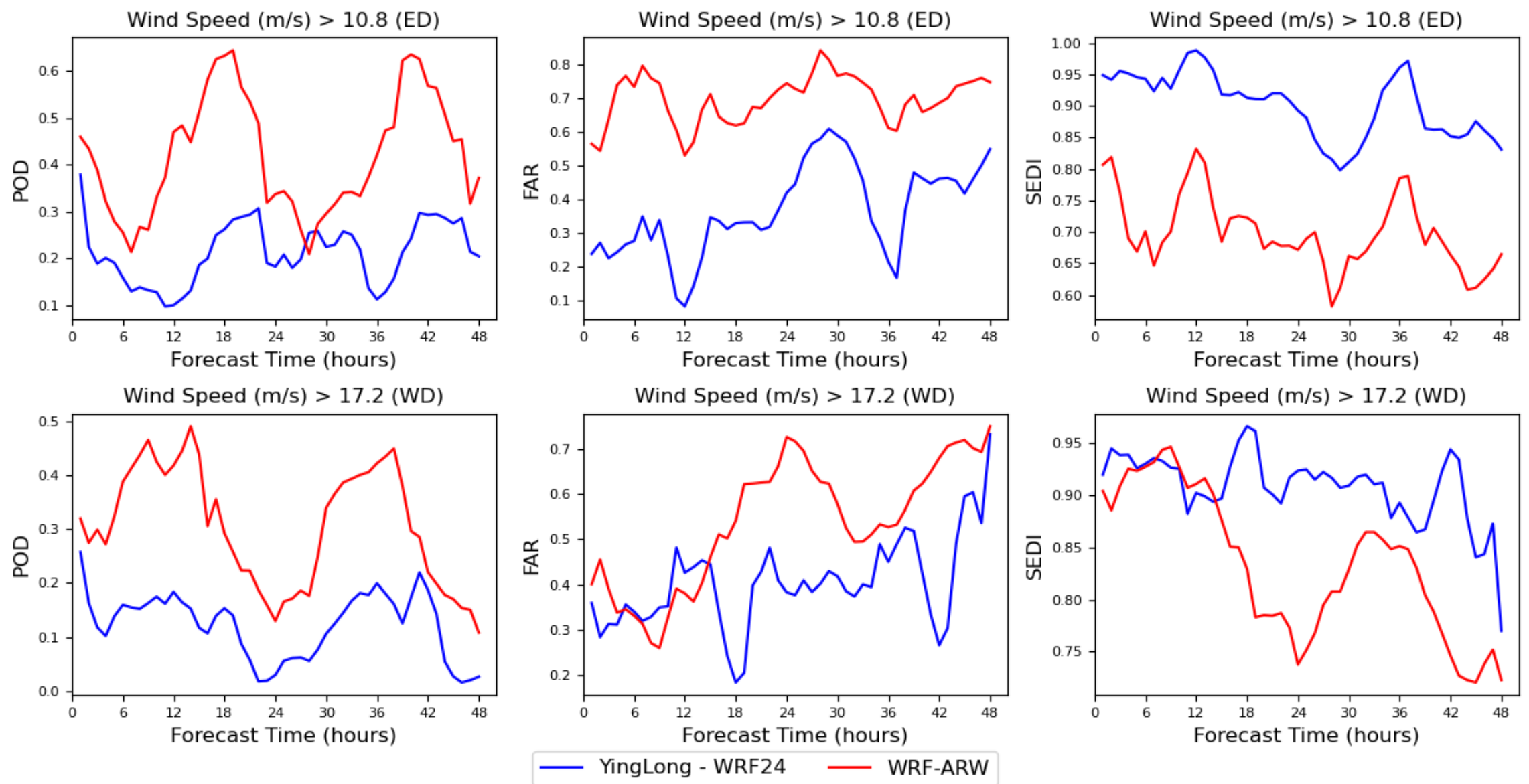

Figure S3. Comparisons of SEDI, POD and FAR for wind speed 10m above surface between YingLong-WRF24 and WRF-ARW. From top to bottom, the extreme wind speed thresholds are 10.8 for ED, and 17.2 m/s for WD, respectively. From left to right, the extreme event metrics are POD (higher is better), FAR (lower is better), and SEDI (higher is better). The blue and red lines represent the corresponding results of WRF-ARW and YingLong, respectively.

**SI 6. ACC results corresponding to Figures 2 and 4**

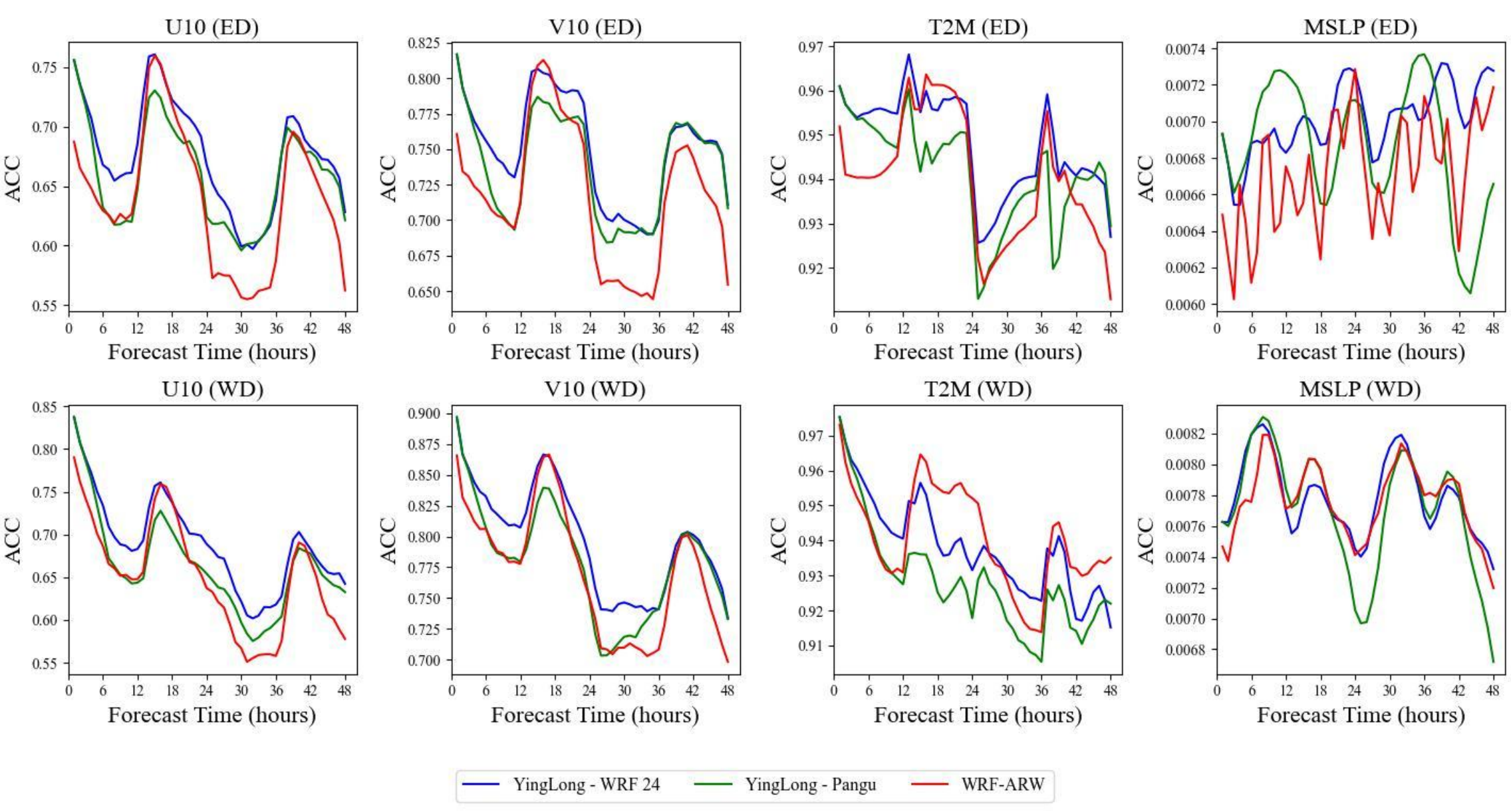

Figure S4. Comparison of ACC of different models based on station observations.

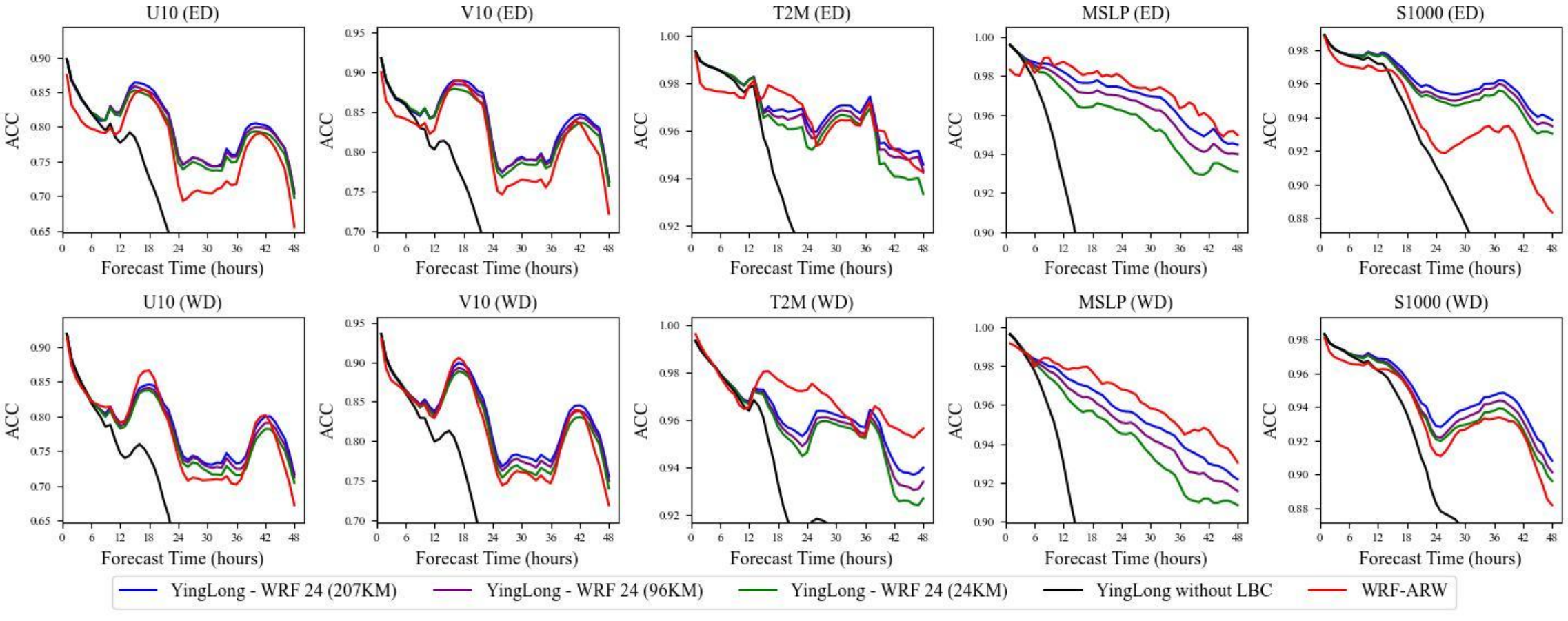

Figure S5. The impact of different regional lateral boundary widths on the inner domain forecast results. The ACC are calculated based on HRRR analysis data. YingLong-WRF24 (207km) means the LBC width for the YingLong model is 207km and is derived from the 24km forecast result of WRF-ARW (which is the YingLong model mainly used in this article). The meanings of other symbols are similar.

**SI 7. Forecast visualizations**

In this part, we select some cases and visualize the corresponding forecast results generated by YingLong and NWP, so that the advantages of YingLong can be found more obviously. Figure S6-S11 show the forecasting results of 10-meter wind speed $\sqrt{\mathrm{U10}^2 + \mathrm{V10}^2}$, 2-meter temperature, and 1000 hPa relative humidity obtained from the YingLong model and NWP model. and 1000 hPa relative humidity obtained from the YingLong model and NWP model utilized in HRRR dataset. It can be found that the results from YingLong model in these cases closely align with the HRRR analysis data. And by calculating the mean absolute error (MAE) for the forecasting results of YingLong and NWP with the analysis data, show the corresponding spatial distribution maps of MAE for each variable, we can further observe that the YingLong forecast results are obviously superior to those of the NWP.

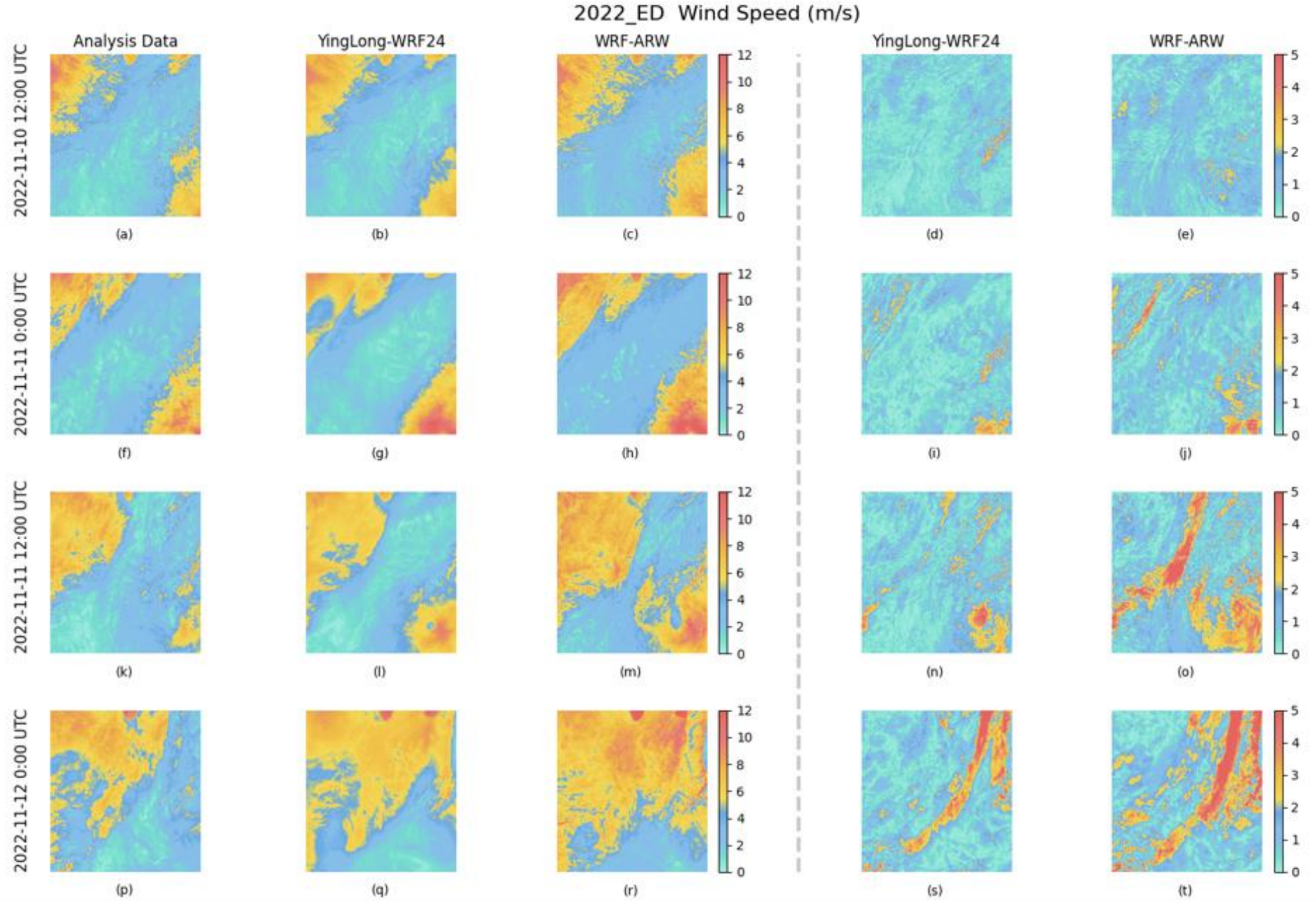

Figure S6. Forecast visualization: wind speed in ED. Forecast initialized at 2022-11-10 00:00 UTC, with plots corresponding to 12, 24, 36 and 48 hour lead times. The three columns on the left are analysis data, YingLong-WRF24, and WRF-ARW visualization results. The two columns on the right are the spatial distribution of MAE of YingLong-WRF24 and WRF-ARW.

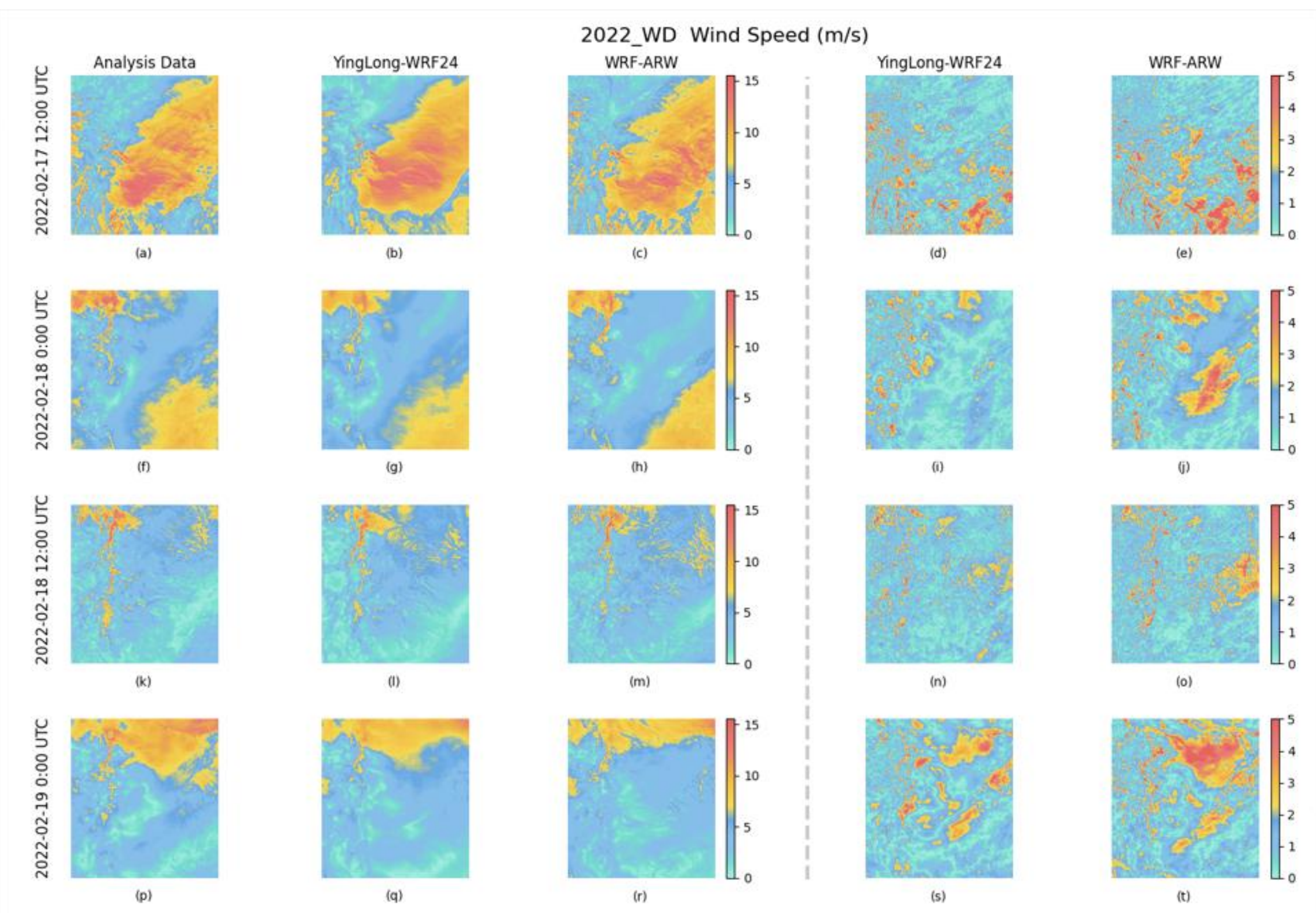

Figure S7. Forecast visualization: wind speed in WD. Forecast initialized at 2022-02-17 00:00 UTC, with plots corresponding to 12, 24, 36 and 48 hour lead times. The three columns on the left are analysis data, YingLong-WRF24, and WRF-ARW visualization results. The two columns on the right are the spatial distribution of MAE of YingLong-WRF24 and WRF-ARW.

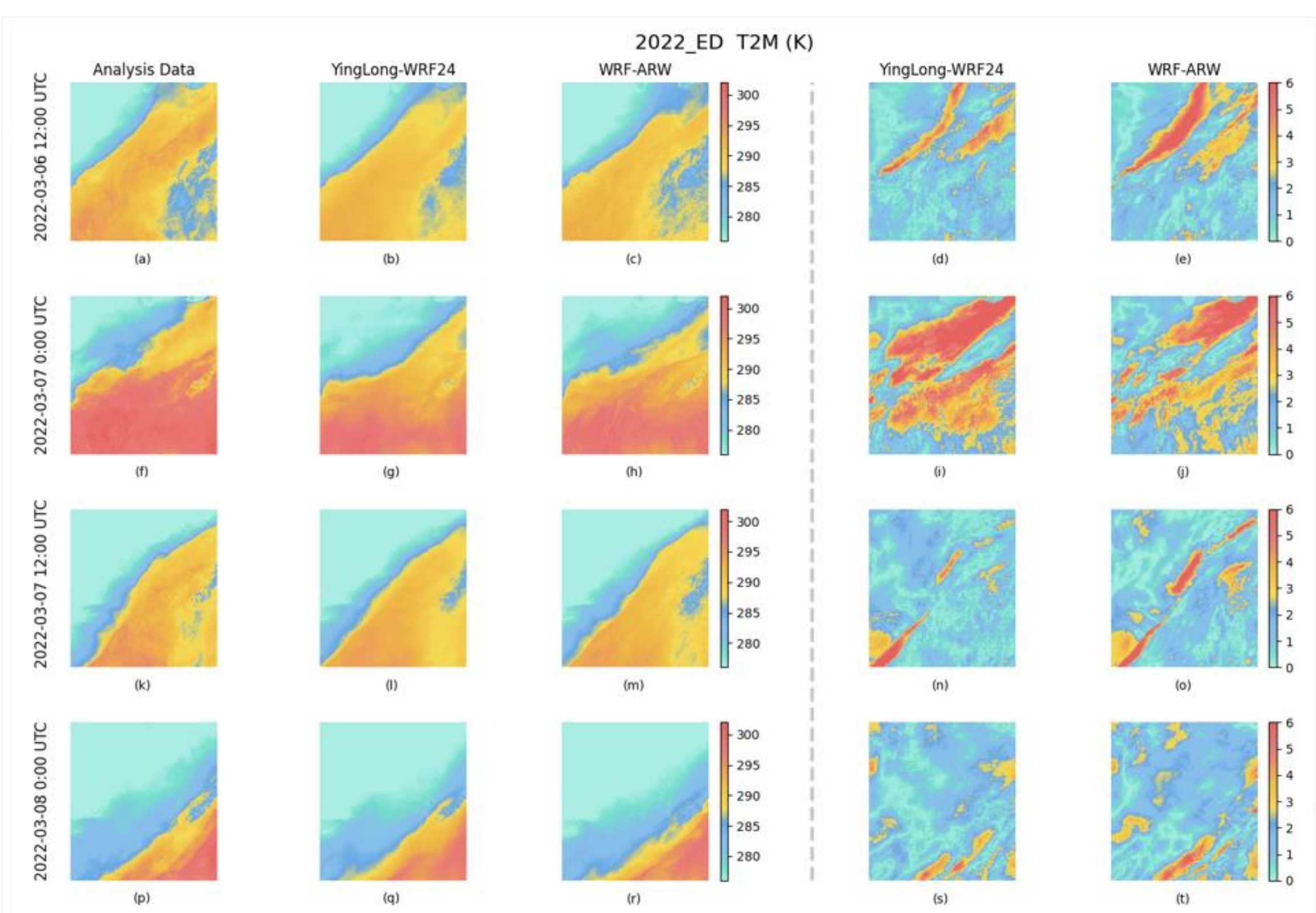

Fig. S8 Forecast visualization: T2M in ED. Forecast initialized at 2022-03-07 00:00 UTC, with plots corresponding to 12, 24, 36 and 48 hour lead times. The three columns on the left are analysis data, YingLong-WRF24, and WRF-ARW visualization results. The two columns on the right are the spatial distribution of MAE of YingLong-WRF24 and WRF-ARW.

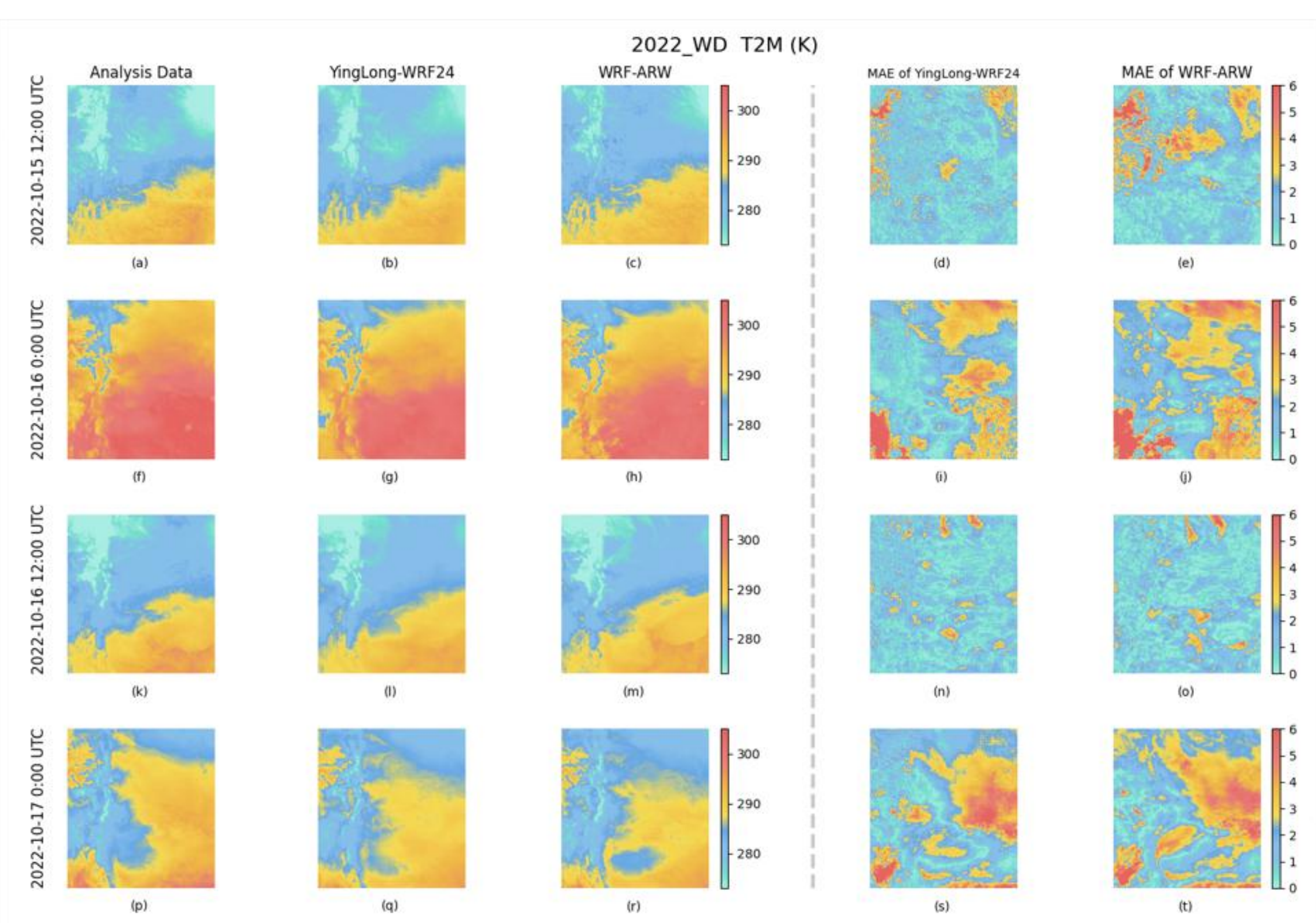

Figure S9. Forecast visualization: T2M in WD. Forecast initialized at 2022-10-15 00:00 UTC, with plots corresponding to 12, 24, 36 and 48 hour lead times. The three columns on the left are analysis data, YingLong-WRF24, and WRF-ARW visualization results. The two columns on the right are the spatial distribution of MAE of YingLong-WRF24 and WRF-ARW.

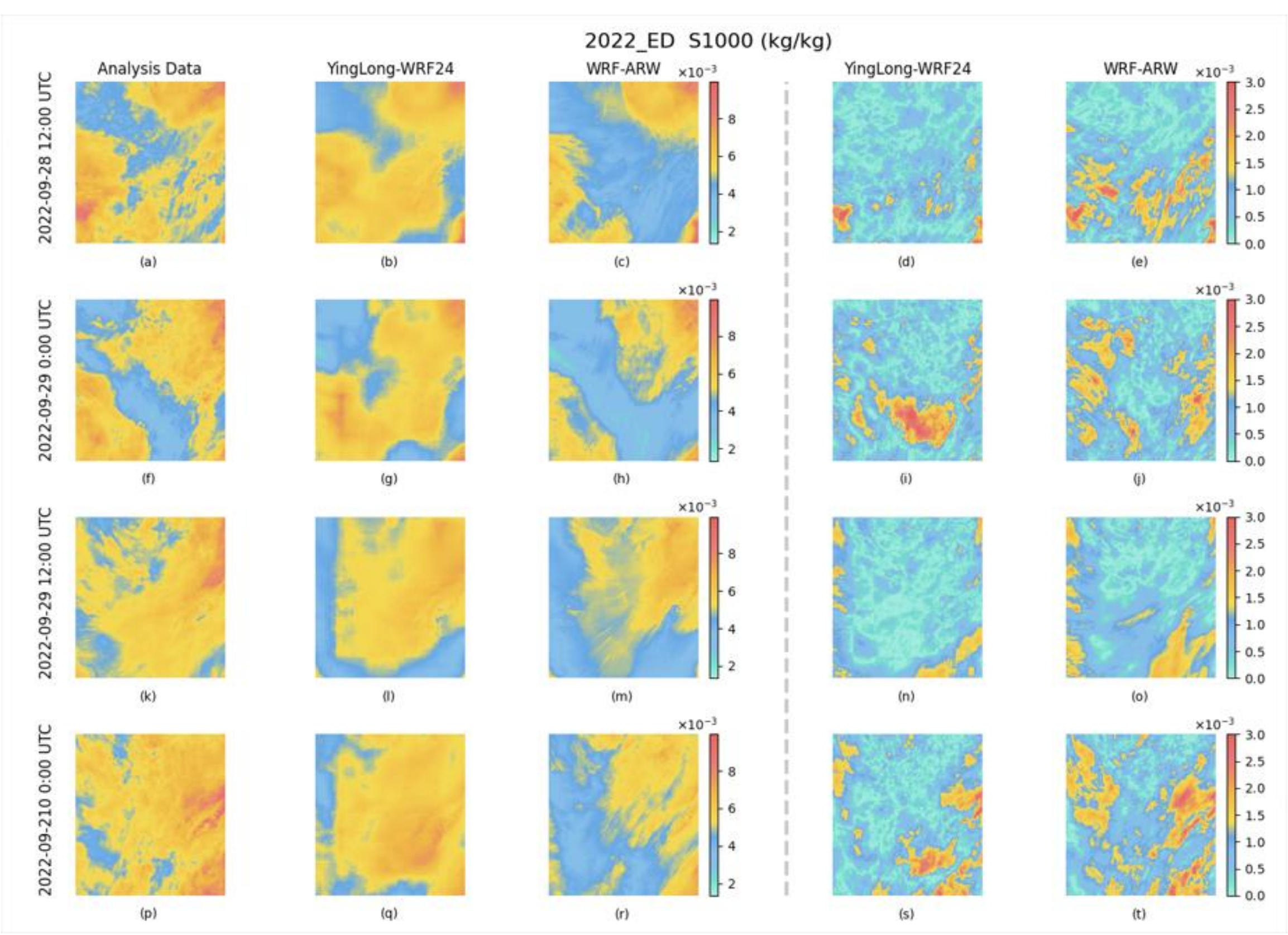

Figure S10. Forecast visualization: S1000 in ED. Forecast initialized at 2022-09-28 00:00 UTC, with plots corresponding to 12, 24, 36 and 48 hour lead times. The three columns on the left are analysis data, YingLong-WRF24, and WRF-ARW visualization results. The two columns on the right are the spatial distribution of MAE of YingLong-WRF24 and WRF-ARW.

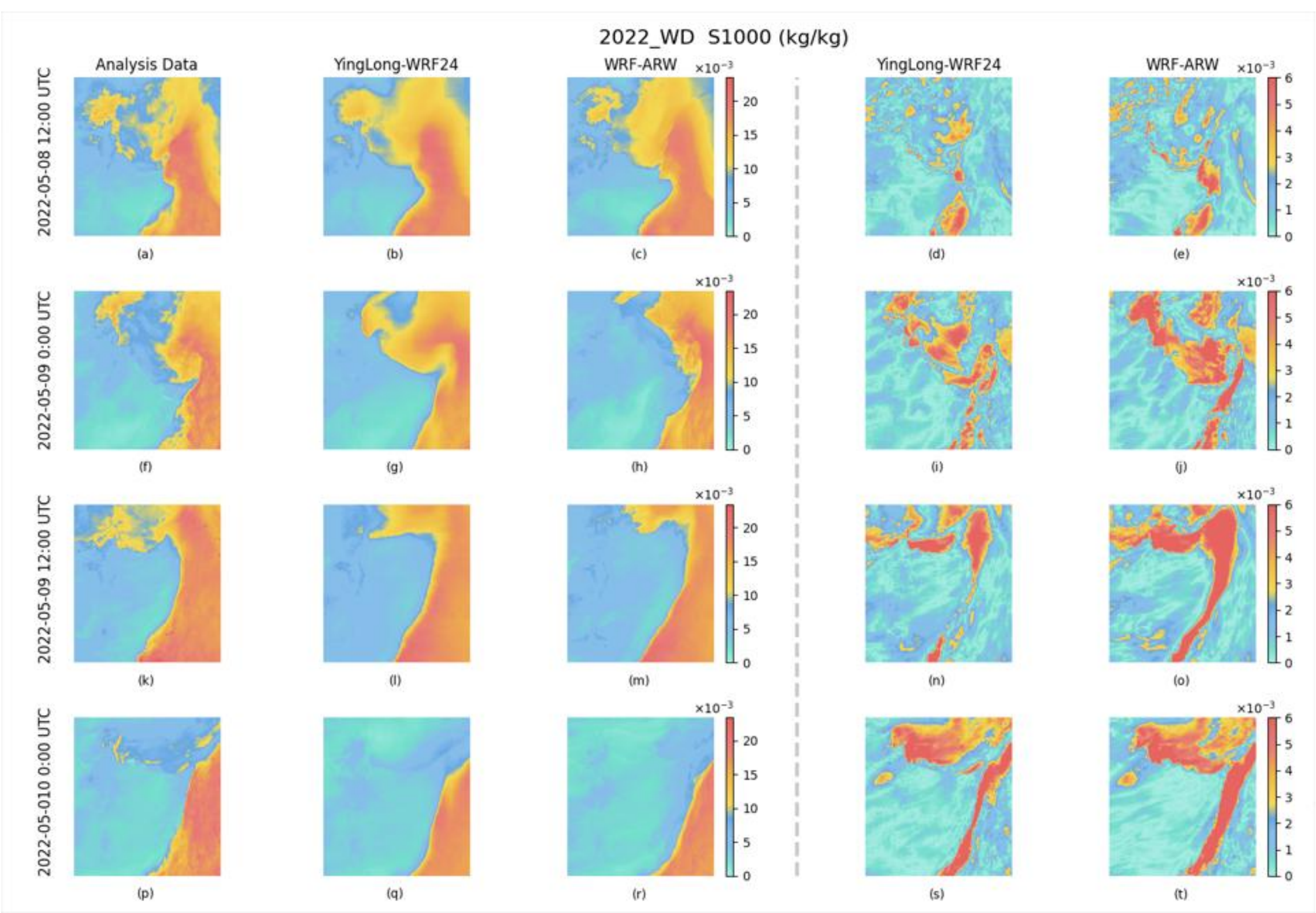

Figure S11. Forecast visualization: S1000 in WD. Forecast initialized at 2022-05-08 00:00 UTC, with plots corresponding to 12, 24, 36 and 48 hour lead times. The three columns on the left are analysis data, YingLong-WRF24, and WRF-ARW visualization results. The two columns on the right are the spatial distribution of MAE of YingLong-WRF24 and WRF-ARW.